\documentclass[letterpaper]{article}

\usepackage[T1]{fontenc}

\usepackage{geometry}
\usepackage{setspace}

\newcommand{\eps}{\varepsilon}
\newcommand{\veps}{\varepsilon_{\mathbf{0}}}
\newcommand{\bfg}{\mathbf{g}}
\newcommand{\bfk}{\mathbf{k}}
\newcommand{\bfr}{\mathbf{r}}
\newcommand{\bfe}{\mathbf{E}}
\newcommand{\sumg}{\sum_{\mathbf{g}}}
\newcommand{\iu}{\mathrm{i}\mkern1mu}
\newcommand{\eu}{\mathrm{e}\mkern1mu}
\newcommand{\opp}[1]{\hat{\mathbf{#1}}} %
\newcommand{\intz}{\int_{-h}^{h} \mathrm{d}z\,}
\newcommand{\intzp}{\int_{-h}^{h} \mathrm{d}z'\,}
\newcommand{\intzpp}{\int_{-h}^{h} \mathrm{d}z''\,}
\newcommand{\uvect}[1]{\hat{\mathbf{e}}_{#1}} 
\usepackage{braket}
\usepackage{verbatim}

\usepackage{graphicx}
\usepackage{float}
\newfloat{scheme}{htbp}{los}
\floatname{scheme}{Scheme}
\floatname{chart}{Chart}
\newfloat{graph}{htbp}{loh}
\usepackage{amsmath,amsfonts,amssymb}
\usepackage[dvipsnames]{xcolor}
\usepackage{authblk}
\author[a,c]{Chi Wang}
\author[a,c]{Dmitrii Gromyko}
\author[b]{Polina Pantiukhina}
\author[a,*]{Lin Wu}
\author[b,*]{Kirill Koshelev}
\affil[a]{Science, Mathematics, and Technology (SMT), Singapore University of Technology and Design (SUTD), 8 Somapah Road,
Singapore 487372}
\affil[b]{Department of Electronic Materials Engineering, Research School of Physics, Australian National University, Canberra, ACT 2601, Australia}
\affil[c]{These authors contributed equally}

\title{Crystal orientation enables near-unity nonlinear circular dichroism in resonant metasurfaces}

\usepackage[unicode,bookmarksnumbered,colorlinks]{hyperref}
\hypersetup{
  allcolors=black,
  pdftitle={Crystal orientation enables near-unity nonlinear circular dichroism in resonant metasurfaces},
  pdfauthor={Chi Wang; Dmitrii Gromyko; Polina Pantiukhina; Lin Wu; Kirill Koshelev},
  bookmarksdepth=0,
  linktoc=all
}
\date{October 3, 2026}
\newcounter{smreference}
\newcommand{\SMbibitem}[1]{%
  \stepcounter{smreference}%
  \bibitem[S\arabic{smreference}]{#1}%
}

\makeatletter
\newcommand{\StartSupportingInformation}{%
  \clearpage
  \newgeometry{left=54.2025pt,top=52.27pt,textwidth=510pt,textheight=672pt,
    headheight=25pt,headsep=16pt,footskip=30pt}%
  \setlength{\paperwidth}{210mm}%
  \setlength{\paperheight}{297mm}%
  \ifdefined\pdfpagewidth
    \pdfpagewidth=\paperwidth \pdfpageheight=\paperheight
  \fi
  \renewcommand{\encodingdefault}{OT1}%
  \renewcommand{\normalsize}{%
    \@setfontsize\normalsize{10}{11.5}%
    \abovedisplayskip 10pt plus 2pt minus 5pt
    \abovedisplayshortskip 0pt plus 3pt
    \belowdisplayshortskip 6pt plus 3pt minus 3pt
    \belowdisplayskip\abovedisplayskip
    \let\@listi\@listI}%
  \renewcommand{\small}{%
    \@setfontsize\small{9}{10.5}%
    \abovedisplayskip 8.5pt plus 3pt minus 4pt
    \abovedisplayshortskip 0pt plus 2pt
    \belowdisplayshortskip 4pt plus 2pt minus 2pt
    \belowdisplayskip\abovedisplayskip}%
  \renewcommand{\footnotesize}{\@setfontsize\footnotesize{8}{9.5}}%
  \normalfont\normalsize
  \setlength{\parindent}{10pt}%
  \setlength{\parskip}{0pt}%
  \setlength{\emergencystretch}{1em}%
  \renewcommand{\footnoterule}{\kern-3pt\hrule width36pt\kern2.6pt}%
  \long\def\@makefntext##1{\parindent\z@\noindent\@makefnmark\space##1}%
  \setlength{\arraycolsep}{3pt}%
  \setlength{\tabcolsep}{2pt}%
  \setlength{\belowcaptionskip}{2pt}%
  \renewcommand{\intz}{\int\limits_{-a}^{a} \mathrm{d}z\,}%
  \renewcommand{\intzp}{\int\limits_{-a}^{a} \mathrm{d}z'\,}%
  \renewcommand{\intzpp}{\int\limits_{-a}^{a} \mathrm{d}z''\,}%
  \setcounter{page}{1}%
  \renewcommand{\thepage}{S\arabic{page}}%
  \setcounter{section}{0}%
  \setcounter{subsection}{0}%
  \setcounter{subsubsection}{0}%
  \setcounter{paragraph}{0}%
  \setcounter{equation}{0}%
  \setcounter{figure}{0}%
  \setcounter{table}{0}%
  \setcounter{footnote}{0}%
  \setcounter{smreference}{0}%
  \setcounter{secnumdepth}{4}%
  \setcounter{tocdepth}{3}%
  \renewcommand{\thesection}{S\arabic{section}}%
  \renewcommand{\thesubsection}{\Alph{subsection}}%
  \renewcommand{\thesubsubsection}{\arabic{subsubsection}}%
  \renewcommand{\theparagraph}{\alph{paragraph}}%
  \renewcommand{\p@subsection}{\thesection}%
  \renewcommand{\p@subsubsection}{\thesection\thesubsection.}%
  \renewcommand{\theequation}{S\arabic{equation}}%
  \renewcommand{\thefigure}{S\arabic{figure}}%
  \renewcommand{\thetable}{S\arabic{table}}%
  \renewcommand{\thefootnote}{\fnsymbol{footnote}}%
  \renewcommand{\theHsection}{SM.\arabic{section}}%
  \renewcommand{\theHsubsection}{SM.\arabic{section}.\arabic{subsection}}%
  \renewcommand{\theHsubsubsection}{SM.\arabic{section}.\arabic{subsection}.\arabic{subsubsection}}%
  \renewcommand{\theHparagraph}{SM.\arabic{section}.\arabic{subsection}.\arabic{subsubsection}.\arabic{paragraph}}%
  \renewcommand{\theHequation}{SM.\arabic{equation}}%
  \renewcommand{\theHfigure}{SM.\arabic{figure}}%
  \renewcommand{\theHtable}{SM.\arabic{table}}%
  \renewcommand{\theHfootnote}{SM.\arabic{footnote}}%
  \hypersetup{linkcolor=RoyalBlue,citecolor=RoyalBlue,
    urlcolor=RoyalBlue,bookmarksdepth=3}%
  \def\ps@supporting{%
    \def\@oddhead{\normalfont\normalsize\hfil\thepage}%
    \let\@evenhead\@oddhead
    \let\@oddfoot\@empty\let\@evenfoot\@empty}%
  \pagestyle{supporting}%
  \def\@seccntformat##1{\csname the##1\endcsname.\quad}%
  \def\SM@sectionstyle{\@startsection{section}{1}{\z@}%
    {0.8cm plus 1ex minus .2ex}{0.5cm}%
    {\normalfont\small\bfseries\centering}}%
  \RenewDocumentCommand\section{s o m}{%
    \IfBooleanTF{##1}%
      {\SM@sectionstyle*{\MakeUppercase{##3}}}%
      {\IfNoValueTF{##2}%
        {\SM@sectionstyle[##3]{\MakeUppercase{##3}}}%
        {\SM@sectionstyle[##2]{\MakeUppercase{##3}}}}}%
  \renewcommand{\subsection}{\@startsection{subsection}{2}{\z@}%
    {0.8cm plus 1ex minus .2ex}{0.5cm}%
    {\normalfont\small\bfseries\centering}}%
  \renewcommand{\subsubsection}{\@startsection{subsubsection}{3}{\z@}%
    {0.8cm plus 1ex minus .2ex}{0.5cm}%
    {\normalfont\small\itshape\centering}}%
  \renewcommand{\paragraph}{\@startsection{paragraph}{4}{\parindent}%
    {0pt}{-1em}{\normalfont\normalsize\itshape}}%
  \renewcommand{\tableofcontents}{\section*{Contents}\@starttoc{toc}}%
  \def\SM@tocline##1##2##3##4{%
    \par\begingroup
      \normalfont\normalsize
      \parindent\z@ \leftskip##1\relax
      \rightskip3em \parfillskip-3em
      \def\numberline####1{\hb@xt@##2{####1.\hfil}}%
      \noindent##3\nobreak\hfill\nobreak
      \hb@xt@3em{\hfil##4}\par
    \endgroup}%
  \def\l@section##1##2{\addvspace{10pt}\SM@tocline{0pt}{1.8em}{##1}{##2}}%
  \def\l@subsection##1##2{\SM@tocline{1.8em}{1.5em}{##1}{##2}}%
  \def\l@subsubsection##1##2{\SM@tocline{3.3em}{1.5em}{##1}{##2}}%
  \let\l@paragraph\@gobbletwo
  \renewcommand{\figurename}{FIG.}%
  \long\def\@makecaption##1##2{%
    \vskip\abovecaptionskip
    \begingroup\small\rmfamily
      \sbox\@tempboxa{##1. ##2}%
      \ifdim\wd\@tempboxa>\hsize
        ##1. ##2\par
      \else
        \global\@minipagefalse
        \hb@xt@\hsize{\hfil\box\@tempboxa\hfil}%
      \fi
    \endgroup
    \vskip\belowcaptionskip}%
  \renewenvironment{thebibliography}[1]{%
    \section*{References}%
    \phantomsection\addcontentsline{toc}{section}{References}%
    \small
    \list{\@biblabel{\@arabic\c@enumiv}}{%
      \settowidth\labelwidth{\@biblabel{##1}}%
      \leftmargin\labelwidth\advance\leftmargin\labelsep
      \usecounter{enumiv}\let\p@enumiv\@empty
      \setlength{\itemsep}{0pt}\setlength{\parsep}{0pt}}%
    \sloppy\clubpenalty4000\widowpenalty4000\sfcode`\.=1000\relax
  }{\endlist}%
  \pdfbookmark[0]{Supporting Information}{supporting-information}%
}
\makeatother

\begin{document}
\pdfbookmark[0]{Main text}{main-text}

\maketitle

\begin{abstract}
Achieving near-unity nonlinear circular dichroism (CD) typically requires geometric symmetry breaking, making it challenging to combine helicity selectivity, resonant enhancement, and structural simplicity. We show that crystal orientation enables near-unity third-harmonic (TH) CD in highly symmetric resonant metasurfaces formed by a square lattice of circular holes in a cubic nonlinear material. Under circularly polarized excitation, strong near-field components of both helicities enabled by the rotational symmetry of the metasurface open two symmetry-allowed in-plane TH generation pathways. Crystal orientation controls their nonlinear coupling, selectively enabling destructive interference for one incident helicity. Our nonlinear temporal coupled-mode theory identifies the conditions for near-unity CD, reveals the distinct roles of pump and TH resonances, and explains the fourfold and eightfold crystal-orientation dependence. Across silicon, germanium, and diamond, we obtain resonantly enhanced TH CD exceeding $99\%$ in the near-infrared, establishing crystal orientation as a design degree of freedom for nonlinear CD in highly symmetric metasurfaces.
\end{abstract}

{*Email: Lin Wu, lin\_wu@sutd.edu.sg,\, Kirill Koshelev, kirill.koshelev@anu.edu.au}

\begin{figure}
\centering
\includegraphics[width=0.8\linewidth]{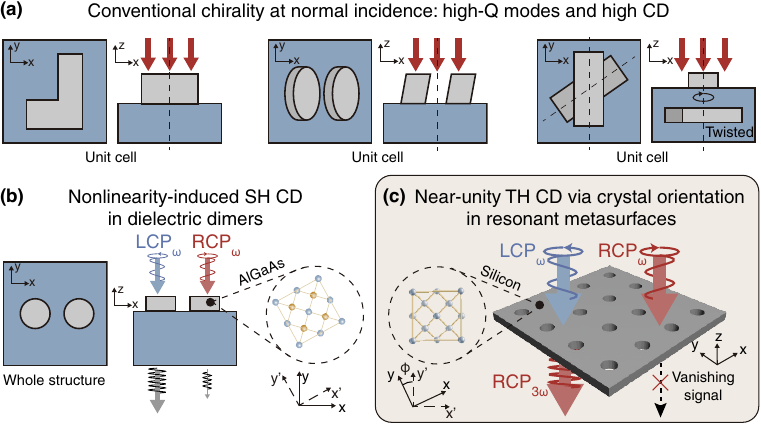}
\caption{Concept of near-unity nonlinear circular dichroism via crystal orientation.
(a) Conventional chiral metasurface approaches for generation of maximal resonant CD in linear scattering and emission: planar geometries with a substrate~\cite{gorkunov2025substrate}, three-dimensional slanted structures~\cite{zhang2022chiral}, and twisted bilayers~\cite{tanaka2020chiral}. (b) Conventional approach of nonlinearity-induced CD of the SH signal in dielectric dimers due to relative rotation of the crystal principal axes~\cite{frizyuk2021nonlinear}. 
(c)  The concept of near-unity TH circular dichroism via crystal orientation in dielectric metasurfaces. A linearly achiral metasurface generates cross-polarized third-harmonic signal with maximal asymmetry under LCP and RCP excitation, yielding near-unitary $\mathrm{CD}_{\mathrm{THG}}$. Inset: Si crystal unit cell rotated relative to the metasurface principal axes by an angle $\phi$.
}
\label{fig:1}
\end{figure}

Chiral metasurfaces provide a versatile platform for precision control of different optical responses to left- and right-circularly polarized  (LCP and RCP) light~\cite{khaliq2023recent, fagiani2023modelling,gorkunov2024rational} manifested in circular dichroism (CD) and optical activity. They find vast applications in chiral spectroscopy~\cite{hu2019high}, spin-selective valleytronics~\cite{li2025valley} and orbital angular momentum control~\cite{ni2021giant,wang2025resonant}, chiral sensing~\cite{solomon2018enantiospecific,both2022nanophotonic}, deciphering light transformations~\cite{tong2026deciphering}, and nonlinear optics~\cite{koshelev2023nonlinear,koshelev2023resonant,tonkaev2025nonlinear}. Figure~\ref{fig:1}(a) summarizes common routes to chirality in linear scattering and emission: substrate-induced symmetry breaking~\cite{nechayev2019substrate,gorkunov2025substrate}, meta-atom rotation~\cite{gryb2023two,toftul2024chiral,sinev2025chirality}, excitation geometry~\cite{papakostas2003optical,ren2012giant}, complex meta-atom shapes~\cite{zhang2022chiral,wang2024enhanced}, twisted bilayers~\cite{tanaka2020chiral,gromyko2024unidirectional}, and material anisotropy~\cite{wang2025chiral, ji2026intrinsically}. 

One widely used route to nonlinear chirality considers nonlinear CD as a consequence of resonantly enhanced linear chirality~\cite{koshelev2023nonlinear}, where LCP and RCP excitation produce different local fields at the fundamental frequency. The nonlinear dependence of second- and third-harmonic generation (SHG and THG) on these fields can amplify the initial contrast, producing strong nonlinear CD. This idea has enabled meta-atom rotation-induced nonlinear chirality~\cite{toftul2024chiral,li2026meta}, chiral high-harmonic generation~\cite{jangid2025chiral}, and nonlinear chiral polaritonics~\cite{heimig2026chiral}. These designs usually break in-plane mirror symmetry to couple incident light to high-$Q$ resonances. Out-of-plane mirror symmetry, however, can be preserved and does not prevent a nonzero nonlinear CD~\cite{koshelev2023nonlinear,koshelev2024scattering,tonkaev2025nonlinear} that can be attributed to the Lorentz reciprocity properties for harmonic generation~\cite{boroviks2023demonstration} and nonlinear asymmetric transmission~\cite{kruk2022asymmetric}.

The harmonic field does not necessarily need to inherit the polarization of the excitation. Instead, near-field engineering can control its handedness, relaxing the geometrical symmetry requirements. Recently, metasurfaces with in-plane mirror symmetries were shown to convert linearly polarized excitation into chiral second-harmonic (SH)~\cite{liu2025dynamic} and third-harmonic (TH)~\cite{lai2025nonlinear} emission. Related approaches were developed to produce circularly polarized TH diffraction beams~\cite{tian2026wavefront} and to convert arbitrary input polarization into circularly polarized harmonic emission~\cite{gromyko2025enabling}.

The response of nonlinear metastructures to chiral light more generally depends on the interplay of symmetries of the meta-atom, lattice, and nonlinear susceptibility tensor~\cite{chen2014symmetry,konishi2014polarization,achouri2023spatial}. A milestone paper by Frizyuk et al.~\cite{frizyuk2021nonlinear} showed that nonzero circular dichroism of the SH signal can be experimentally achieved in AlGaAs disk dimers where the crystal orientation is rotated with respect to the dimer axes, schematically shown in Fig.~\ref{fig:1}(b). This concept was subsequently extended to SH CD in other achiral nanoparticles~\cite{nikitina2023nonlinear,nikitina2024achiral} and  metasurfaces~\cite{menshikov2026tailoring} made of GaAs and BaTiO$_3$. The key physics of this approach relies on interference between distinct in-plane and out-of-plane SHG modal pathways whose amplitudes and phases depend on the nanostructure symmetry and crystal orientation. The use of nonlinearity as an independent degree of freedom was highlighted in recent works on nonlinearity-induced torque~\cite{toftul2023nonlinearity}, angular momentum light structuring~\cite{menshikov2025light}, near-infrared polarimetric imaging~\cite{menshikov2026near}, and tensor-driven geometric phase~\cite{guercio2026tensor}.

For scalable CMOS-compatible practical devices, a large CD value must occur at the resonance where harmonic generation is strongest. Approaches based on linear chirality can achieve large nonlinear CD~\cite{koshelev2023nonlinear,toftul2024chiral,tonkaev2025nonlinear}, but often require precise parameter control and symmetry breaking or complex geometries that complicate fabrication and scaling. Crystal-orientation approaches preserve high-symmetry geometries~\cite{frizyuk2021nonlinear,nikitina2024achiral,menshikov2026tailoring}, yet their reported CD at resonance does not exceed $25\%$, while low modal quality factors limit SHG enhancement. The challenge is therefore to combine near-unity CD with the regime of the strongest resonant enhancement while retaining the geometrical simplicity of the metastructure.

Here, we show that the control of crystal orientation in nonlinear metasurfaces with third-order susceptibility allows us to meet these requirements in a simple metasurface design metasurface  shown in Fig.~\ref{fig:1}(c). The design consists of a suspended membrane perforated by a square lattice of circular holes, preserving mirror symmetries in the linear scattering problem about all three coordinate planes. We consider cubic nonlinear crystals, taking crystalline silicon as the main example. Using quasinormal-mode expansion and nonlinear temporal coupled-mode theory, we show that the structure supports high-$Q$ quasi-guided resonances and remains achiral in linear scattering. Yet rotating the nonlinear susceptibility tensor relative to the metasurface axes is sufficient to produce drastically different THG responses to LCP and RCP excitation. Using the local selection rules based on Neumann's principle, we show that a purely circularly polarized incident wave can generate LCP and RCP near-field components of equal magnitude, in contrast to conventional far-field selection rules. These components drive interfering in-plane THG pathways, suppressing one circular emission channel at selected crystal orientations. Unlike earlier schemes~\cite{frizyuk2021nonlinear,menshikov2026tailoring}, this mechanism requires no out-of-plane pathway and is therefore not restricted to materials supporting both contributions. We derive a simple relation between nonlinear CD and crystal orientation, including a transition from fourfold to eightfold angular periodicity when additional resonances are excited at the harmonic frequency. Extending the analysis from Si to Ge, diamond, Cu$_2$O, and CaF$_2$, we find resonant TH CD above $99\%$ across different near-infrared bands. In silicon, this yields a simple, scalable, CMOS-compatible metasurface in which near-unity nonlinear CD coincides with maximal resonant enhancement, without introducing geometrical chirality. 



\begin{figure}
\centering
\includegraphics[width=0.99\linewidth]{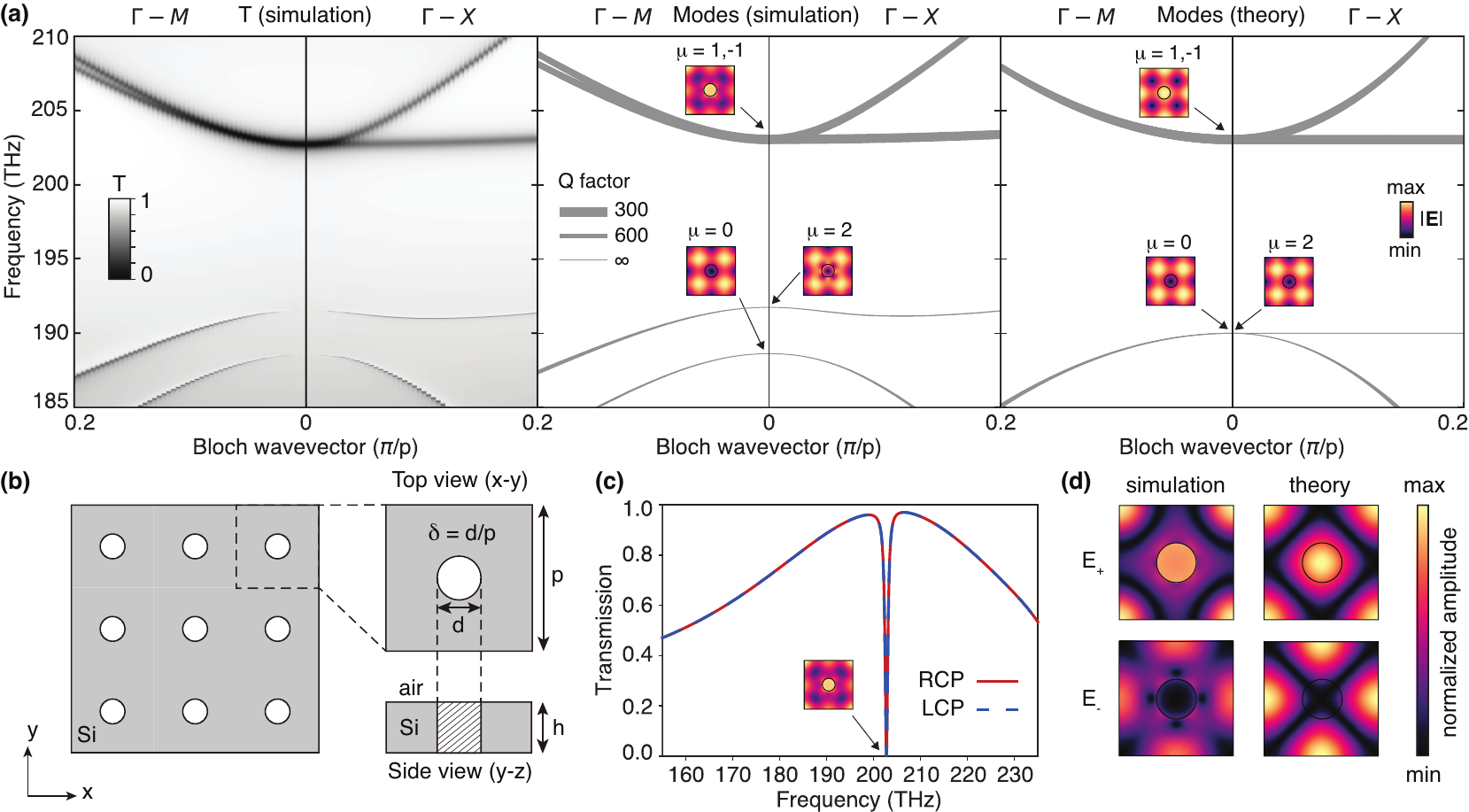}
\caption{{\bf Resonant response at the pump frequency.} (a) Transmittance map (left) and eigenmode spectra (center and right) along the $\Gamma-M$ and $\Gamma-X$ directions for RCP ($\sigma=1$) excitation. The refractive index is $3.5$, $p=577~\mathrm{nm}$, $h=2a=220~\mathrm{nm}$, and $\delta=d/p=0.35$.  In the mode spectrum panels, the plotted curve thickness encodes the inverse quality factor. The right panel shows data fitted to the analytical mode dispersion obtained from the analytical TCMT with parameters listed in the text, see SI, Section \ref{sm:sec:dispersion}. Insets show $|\mathbf{E}|$ for multipolar modes at normal incidence. (b) Metasurface schematic. (c) Transmission spectra for RCP/LCP ($\sigma=\pm 1$) excitation under normal incidence. (d) Co- and cross-polarized near-field components for resonant RCP excitation at $203$~THz, comparison between simulation and theory in Eq.~\eqref{eq:3}. 
}
\label{fig:2}
\end{figure}

We first describe the resonant response of the achiral metasurface schematically shown in Fig.~\ref{fig:2}(b) at the pump frequency. The metasurface is composed of a square lattice of air holes in a crystalline silicon slab suspended in air. The metasurface modal spectrum can be classified based on its point group symmetry at the $\Gamma$ point~\cite{overvig2020selection}. For $C_{4v}$ symmetry, the modes can be characterized by a discrete rotational-symmetry index $\mu=0,\pm 1,2$ that labels the transformation of the modes under fourfold rotation about the metasurface normal. We consider the frequency range in the vicinity of the complex frequency $k_{\mu}$ of four fundamental quasi-guided modes: monopole ($\mu=0$) and quadrupole ($\mu=2$) singlets, and a dipolar doublet ($\mu=\pm 1$), where $k=\omega/c$.

We employ temporal coupled-mode theory (TCMT) to describe the scattering dynamics of the metasurface~\cite{fan2003temporal,ruan2012temporal}. We assume the metasurface is excited with a circularly polarized plane wave incident normal to the metasurface plane. Assuming a harmonic time dependence of the fields proportional to $\eu^{-\iu \omega t}$, we can define the amplitudes of the incoming and outgoing waves $s^{\rm (in)}_{n}(k)$ and $s^{\rm (out)}_{n}(k)$, respectively, with the index $n=\{u,\sigma\}$ that labels the propagation direction $u$ (incidence from the top or bottom) and the helicity $\sigma =\pm 1$. The temporal dynamics of scattering channels and resonant amplitudes ${a}_{\mu}(k)$ can then be written as a set of two coupled TCMT equations~\cite{fan2003temporal,ruan2012temporal}:
\begin{align}
&-\iu k{a}_{\mu}(k) = -\iu k_{\mu} {a}_{\mu}(k) + \sum_{n} {D}_{n,\mu} s^{\rm (in)}_{n}(k),
\label{eq:1} \\
&s^{\rm (out)}_{n'}(k)=\sum_{n}C_{n',n}(k)s^{\rm (in)}_{n}(k)+ \sum_{\mu} {K}_{n',\mu}
{a}_{\mu}(k).
\label{eq:2}
\end{align}
Here, $C_{n',n}(k)$ is a matrix of the direct (non-resonant) scattering process, and ${D}_{n,\mu}$, ${K}_{n,\mu}$ are coupling matrices that describe resonant mode coupling to the corresponding scattering channels.

We can show that the TCMT Eqs.~(\ref{eq:1},\,\ref{eq:2}) can be derived analytically from Maxwell's equations within the three-dimensional spatial coupled-mode approach~\cite{liang2011three}, see Supporting Information, Section \ref{sm:sec:linear}. In this approach,  the formation of quasi-guided modes in a metasurface with periodic permittivity profile is described through the coupling of guided modes, which are waveguide modes of an effective dielectric slab with the in-plane averaged permittivity via periodic modulation. The resulting TCMT coupling matrices can be expressed as ${D}_{n,\mu}=-2\iu \delta_{\mu,\sigma}\sigma  D_{u} $ and ${K}_{n,\mu}=2\iu \delta_{\mu,\sigma}\sigma  D_{u}$, where ${D}_u$ is determined by the overlap of the guided mode profile and internal scattering channel fields in the normal direction. The direct scattering matrix $C_{\sigma',\sigma,u',u}$ is diagonal with respect to $\sigma$ and is defined by the Fresnel reflection and transmission amplitudes of the effective slab. Additionally, the model predicts the mode dispersion in the vicinity of the $\Gamma$ point by treating the in-plane Bloch wavevector as the perturbation parameter; see Supporting Information, Section \ref{sm:sec:dispersion} for more details.


Figure~\ref{fig:2}(a) shows the calculated transmittance and eigenmode spectrum of s-polarized modes of the Si metasurface with thickness $h=2a$ along the high-symmetry directions of the Brillouin zone. We consider the refractive index of $3.5$ with weak frequency dispersion in the near-infrared range~\cite{franta2018determination}.  We select the target metasurface design with the size parameter $\delta=d/p=0.35$ that corresponds to the dipolar mode frequency of $203$~THz and the quality factor \(Q\) of $300$ at the $\Gamma$ point. The transmittance in the left panel is calculated using the open-source rigorous coupled-wave analysis (RCWA) package {\it Inkstone}~\cite{alexysong_inkstone} using a basis of $441$ Fourier harmonics. The mode spectrum in the central panel is calculated using the eigenmode solver in COMSOL Multiphysics. The mode data in the right panel are fitted to the analytical mode dispersion obtained in SI with $k_g=4.11~\mathrm{\mu m}^{-1}$, $\kappa_{2}=0.14~\mathrm{\mu m}^{-1}$, $\gamma=0.0038~\mathrm{\mu m}^{-1}$, $N=0.01~\mathrm{\mu m}$. The spectra show that the monopole and quadrupole modes at the $\Gamma$ point are non-radiative states with degenerate frequencies, and dipolar modes are doubly degenerate radiative states. The model predicts the correct dispersion form and is in good agreement with the numerical results. In the numerical data, the degeneracy between monopole and quadrupole modes is lifted by corrections to the derived model arising from a non-vanishing size parameter.

Figure~\ref{fig:2}(c) shows the transmittance spectrum for RCP/LCP ($\sigma=\pm 1$) excitation for $\delta = 0.35$. Only the $\mu=\pm 1$ modes are visible in the spectrum in the vicinity of $203$~THz due to the polarization selection rule $\mu=\sigma$ imposed by ${D}_{n,\mu}$ and ${K}_{n,\mu}$. In the resonant regime, the corresponding resonant mode amplitude is enhanced because of $k\approx \operatorname{Re}[k_\sigma]$ in Eq.~\eqref{eq:1}. Thus, the total electric field is dominated by the dipolar resonant term $\bfe(\bfr;k)\approx {a}_{\sigma}(k) \bfe_{\sigma}(\bfr)$. The corresponding field $\bfe_{\sigma}(\bfr)$ of the dipolar quasi-normal mode can be explicitly evaluated within the developed spatial coupled-mode theory; see Supporting Information, Section \ref{sm:sec:nearfields} for more details. The resulting co- and cross-polarized components of the near-field, $E_{\pm}(\bfr)=\uvect{\mp\sigma}\cdot\bfe(\bfr)$ for $\uvect{\sigma}=(\uvect{x}-\iu\sigma \uvect{y})/\sqrt{2}$ are 
\begin{equation}
E_{\pm}(\bfr;k)\approx
\frac{\iu\sigma {a}_{\sigma}(k)}{\sqrt{2}}\theta_{g}(z)
\left[
\cos(gy)\pm\cos(gx)
\right].
\label{eq:3}
\end{equation}
We note that the out-of-plane field component $E_z(\bfr)=0$ for the selected basis of $\rm s$-polarized guided modes. Figure~\ref{fig:2}(d) shows the comparison of calculated and theoretical $E_{\pm}(x,y)$ given by Eq.~\eqref{eq:3} at $z=0$ (central plane of the metasurface) at the dipolar mode frequency of $203$~THz. The maximal amplitudes of co- and cross-polarized near-field components are equal. 

\begin{figure}
\centering
\includegraphics[width=0.99\linewidth]{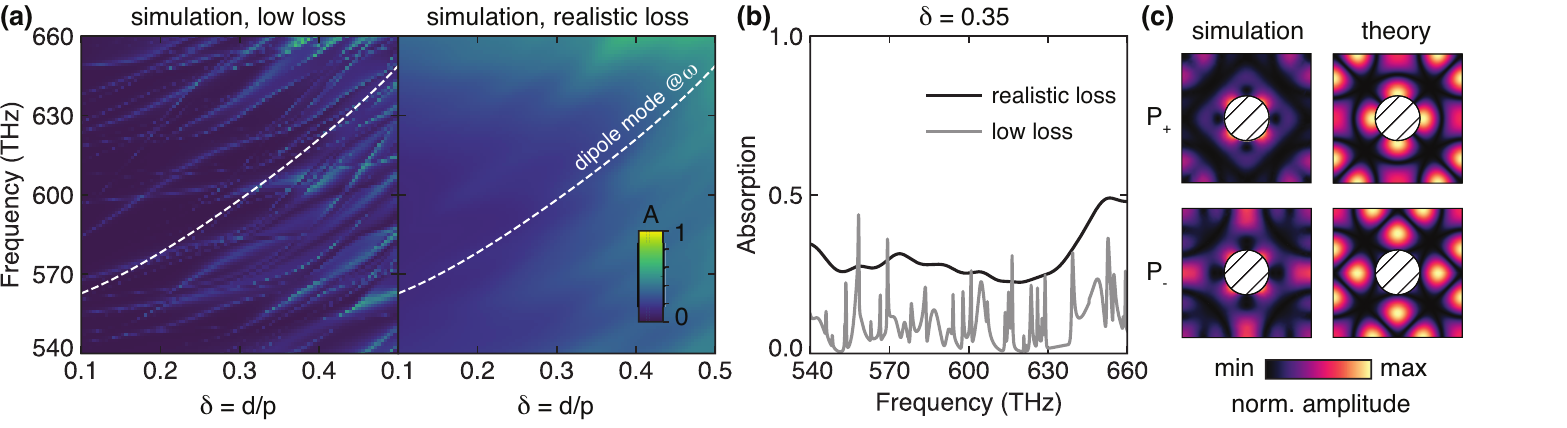}
\caption{{\bf Resonant response  at the TH frequency and nonlinear polarization structure.} 
(a) Absorptance map vs. frequency and size parameter $\delta$  of the metasurface with the refractive index of $4.1$ and extinction coefficient $4\times10^{-4}$ (left, low loss) and $4\times10^{-2}$ (right, realistic loss) for RCP ($\sigma=1$) excitation at the TH frequency under normal incidence. The white dashed line shows the tripled-frequency dispersion of the dipolar mode at the pump frequency. (b) Absorptance spectra for realistic and low-loss cases for $\delta=0.35$. (c) Co- and cross-polarized components of the nonlinear polarization for resonant RCP excitation at $203$~THz and $\phi=0^\circ$, comparison of simulation and theory in Eq.~\eqref{eq:4}.
}
\label{fig:3}
\end{figure}

We further study the resonant response of the metasurface at the third-harmonic (TH) frequency under normal incidence. We use the RCWA package to calculate the absorptance spectra, comparing realistic Si dispersion at $600$ THz~\cite{franta2018determination} with refractive index $n=4.1$ and extinction coefficient $k=4\times10^{-2}$ to an otherwise identical low-loss material with the same refractive index and a reduced extinction coefficient of $k=4\times10^{-4}$.
The other simulation parameters are the same as in Fig.~\ref{fig:2}. Figure~\ref{fig:3}(a) shows the absorptance maps vs. frequency and $\delta$ for both cases. The overlaid white dashed line shows three times the frequency dispersion of the dipolar $|\mu|=1$ modes obtained from eigenmode analysis.  The low-loss map demonstrates a large number of resonant modes that overlap with the tripled dipolar mode frequency, with the highest overlap around $\delta\approx 0.35$. The corresponding TH modes are smeared out in the map for realistic material parameters; however, the broad spectral features are preserved. Figure~\ref{fig:3}(b) shows a cross-section of the absorptance maps at $\delta=0.35$ corresponding to the target metasurface design used in Fig.~\ref{fig:2}.

We next analyze the nonlinear resonant response of the metasurface and demonstrate that it can be described with a nonlinear temporal coupled-mode theory. For crystalline Si, the nonlinear polarization $\mathbf{P}(\bfr;3k)$ can be described by two independent components of the nonlinear susceptibility, $\chi^{(3)}_{1111}$ and $\chi^{(3)}_{1212}$~\cite{moss1989dispersion}. We assume that \(xy\) coordinate frame is aligned with the metasurface axes, and the crystallographic axes of Si are rotated by an angle $\phi$ clockwise as shown in Fig.~\ref{fig:1}(c). Importantly, the nonlinear response of the metasurface is characterized by $C_{4h}$ symmetry; however, the linear metasurface response remains unchanged, conforming to $D_{4h}$ symmetry regardless of the rotation angle. Then, the co- and cross-polarized components of the nonlinear polarization $P_{\pm}=\uvect{\mp\sigma}\cdot\mathbf{P}(\bfr)$ can be expressed via $E_{\pm}(\bfr;k)$ given by Eq.~\eqref{eq:3}, see also Supporting Information, Section \ref{sm:sec:selection},
\begin{equation}
P_{\mp}(\bfr;3k)
=
\varepsilon_0
\left[
\eu^{\pm 4\iu\sigma\phi}\Delta\chi^{(3)}(\bfr)
E_{\pm}^{3}(\bfr;k)+3\chi^{(3)}_{0}(\bfr) E_{\pm}(\bfr;k)E_\mp^{2}(\bfr;k)
\right].
\label{eq:4}
\end{equation}
Here, we omit the terms generated by the out-of-plane field component $E_z(\mathbf{r})$ that is negligible for dipolar modes formed by purely $\rm s$-polarized guided modes. We define two independent components of the nonlinear susceptibility in the helical basis as $\Delta\chi^{(3)}=(\chi^{(3)}_{1111}-3\chi^{(3)}_{1212})/2$ and $\chi^{(3)}_{0}=(\chi^{(3)}_{1111}+\chi^{(3)}_{1212})/2$. We note that inside the air holes $\Delta\chi^{(3)}(\bfr)=\chi^{(3)}_{0}(\bfr)=0$.  Figure~\ref{fig:3}(c) shows the comparison of calculated and theoretical $P_{\pm}(x,y)$  given by Eq.~\eqref{eq:4} at $z=0$ (central plane of the metasurface) at the pump frequency of $203$~THz. For the calculation, we used the anisotropy factor of the nonlinear susceptibility $3\chi^{(3)}_{1212}/\chi^{(3)}_{1111}=1.41$ measured at $192$~THz pump, assuming its weak frequency dispersion in the near-IR range~\cite{moss1989dispersion}. We note that both harmonic pathways in Eq.~\eqref{eq:4} are generated by the in-plane field components $E_{\pm}$, unlike the crystal-orientation induced SH CD in previous demonstrations~\cite{frizyuk2021nonlinear,nikitina2024achiral,menshikov2026tailoring} that required a pathway driven by an out-of-plane component $E_z$ with the corresponding large out-of-plane susceptibility. We also note that the magnitude of $E_{\pm}$, and consequently, of both in-plane TH pathways are comparable due to the modal structure specific to dipolar modes of metasurfaces with a third- or higher-order rotational symmetry, as can be seen from Eq.~\eqref{eq:3}.

Equation~\eqref{eq:4} establishes the local selection rules in the ciruclar basis~\cite{menshikov2026tailoring}, following from Neumann's principle~\cite{konishi2014polarization}, that connect the helicity $\sigma_4$ of the nonlinear polarization components to the helicities of the three near-field components at the pump frequency $\sigma_1,\sigma_2,\sigma_3$,
\begin{equation}
\sigma_1+\sigma_2+\sigma_3-\sigma_4 = 4m.
\label{eq:5}
\end{equation}
Here, $m=0,\pm 1$ are the quanta of the angular momentum projection added by the $C_{4v}$ metasurface lattice that define the relative rotation angle dependence of the phase factors $\eu^{4\iu m\phi}$ in Eq.~\eqref{eq:4}. The solutions of Eq.~\eqref{eq:5} are $\sigma_1=\sigma_2=\sigma_3=-\sigma_4$ for $|m|=1$, and $\sigma_1=\sigma_2=-\sigma_3=\sigma_4$ (plus all permutations) for $m=0$. Due to their different dependence on $\phi$, these allowed pathways enable an interference mechanism that can suppress nonlinear harmonic generation for a selected input helicity. Unlike the THG process considered here, the analogous chiral SHG process from a $C_{3v}$ metasurface does not support such interference, because its helicity phase-matching condition, $\sigma_1+\sigma_2-\sigma_3 = 3m$, allows only a single solution with $|m|=1$. A nonzero SHG CD in $C_{3v}$ metasurfaces thus requires strong contribution of $E_z(\mathbf{r})$ to Eq.~\eqref{eq:5}~\cite{guercio2026tensor}.


We next establish how the suppression of the nonlinear polarization by interference between the two allowed pathways can be identified in the emitted third-harmonic signal. To this end, we develop a nonlinear temporal coupled-mode theory that connects the input scattering channels $s^{\rm (in)}_{n}$ at the pump frequency to the output channels $s^{\rm (out)}_{n}$ at the TH frequency, as well as to the resonant amplitudes ${b}_{\mu}$ of quasi-normal modes that can be additionally excited at the TH frequency due to their high density, as shown in Fig.~\ref{fig:3}(a). In general, the harmonic output depends on the propagation direction and polarization of each input beam, making the problem multidimensional. However, recent studies have introduced an effective nonlinear scattering matrix that directly connects $s^{\rm (out)}_{n}(3k)$ and $s^{\rm (in)}_{n}(k)$ for the special case of degenerate sum-frequency generation driven by a chiral input beam~\cite{koshelev2024scattering}. Following this approach and using the analytical model established for the derivation of Eqs.~(\ref{eq:1},\,\ref{eq:2}), we show in Supporting Information, Section \ref{sm:subsec:nonlinear-derivation} that the nonlinear TCMT can be written as
\begin{equation}
\begin{aligned}
&-3\iu k {b}_{\mu}(3k)=-\iu{k}^{(3)}_{\mu}
{b}_{\mu}(3k)+  \sum_n{D}^{(3)}_{n,\mu}[s^{\rm (in)}_{n}(k)]^3,\\
&s^{\rm (out)}_{n'}(3k)=\sum_n C^{(3)}_{n',n}(k)[s^{\rm (in)}_{n}(k)]^3+\sum_\mu  {K}^{(3)}_{n',\mu} b_{\mu}(3k),
\end{aligned}    
\label{eq:6}
\end{equation}
Here, ${k}^{(3)}_{\mu}$ are the complex mode frequencies of the TH modes with $\mu=0,\pm 1,2$. The outcoupling matrix is ${K}^{(3)}_{n,\mu}=2\iu \delta_{\mu,\sigma}\sigma  D^{(3)}_{u}$, where $D^{(3)}_{u}$ is defined at the TH frequency similar to $D_{u}$. The nonlinear coupling matrix can be calculated as
\begin{equation}
{D}^{(3)}_{n,\mu}\propto\int {\rm d}\bfr \left[E^{(3)}_{\mu,+}(\bfr)\right]^*P_{+}(\bfr)+\int {\rm d}\bfr\left[E^{(3)}_{\mu,-}(\bfr)\right]^*P_{-}(\bfr),   
\label{eq:7}
\end{equation}
where $E^{(3)}_{\mu,\pm}(\bfr)$ are co- and cross-polarized components of the TH mode fields. The direct nonlinear scattering matrix is $C^{(3)}_{\sigma',\sigma,u',u}(k)$ converts input helicity $\sigma$ to the output helicity $\sigma'=-\sigma$. We note that cross-terms such as $[s^{\rm (in)}_{n}]^2s^{\rm (in)}_{n'}$ do not appear in Eq.~\eqref{eq:6} because for the degenerate THG process we consider a single-channel excitation where the polarization and direction of three input beams are identical.

\begin{figure}
\centering
\includegraphics[width=0.95\linewidth]{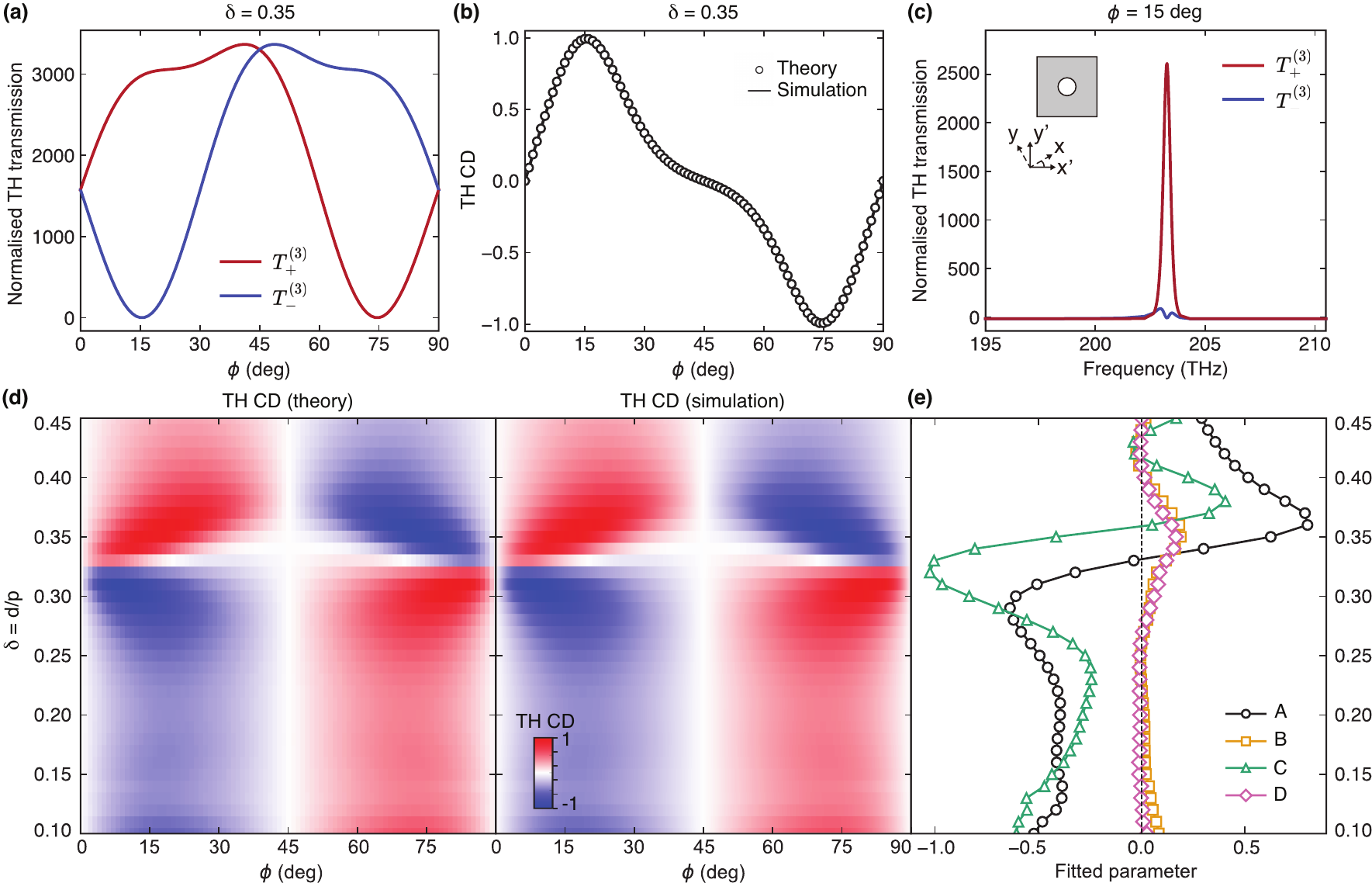}
\caption{{\bf Nonlinear resonant CD controlled by crystal orientation.}
(a, b) Calculated (lines) and fitted (markers) TH transmission (a) and TH CD (b) vs. the relative angle $\phi$ between the metasurface and $\hat{\chi}^{(3)}$ in-plane axes for the target metasurface design with $\delta=0.35$. The pump frequency is $203$~THz, with RCP ($\sigma=1$) and LCP ($\sigma=-1$) excitation. The theoretical curve is fitted to Eq.~\eqref{eq:9}. (c) Pump frequency spectrum of the normalized TH transmission for the optimal angle $\phi=15^\circ$.
(d) Calculated and fitted theoretical TH CD in the $(\phi,\delta)$ parameter space. The data are evaluated for the resonance frequency of the pump dipole mode for each value of $\delta$, following the white dashed curve in Fig.~\ref{fig:3}(a).
(e) Fitted parameters $A$, $B$, $C$ and $D$ in Eq.~\eqref{eq:9} used for the TH CD map in (d). 
}
\label{fig:4}
\end{figure}

Using Eq.~\eqref{eq:6} with the exact expressions for the coupling matrices, we evaluate the effective nonlinear scattering matrix defined as $s^{\rm (out)}_{n'}(3k)=\sum_n S^{(3)}_{n',n}(k)[s^{\rm (in)}_{n}(k)]^3$, see Supporting Information, Section \ref{sm:subsec:nonlinear-scattering}. Its nonzero components are defined by the far-field polarization selection rule $\sigma'=-\sigma$. The TH transmittance in the zeroth-order diffraction channel for RCP and LCP excitation is $T^{(3)}_{\pm}=|S^{(3)}_{\mp1,{\rm b},\pm1,{\rm t}}|^2$ and can be evaluated as
\begin{equation}
T^{(3)}_{\pm}\propto\left|I_{1,z}\left[\eu^{\pm 4\iu\phi}\Delta\chi^{(3)}+3\chi^{(3)}_{0}\right]+\frac{ I_{2,z}}{(3k-k^{(3)}_{\mu=1})}\left[\cos{(4\phi)}\Delta\chi^{(3)}+3\chi^{(3)}_{0}\right]\right|^2.
\label{eq:8}
\end{equation}
Here, $I_{1,z}$ and $I_{2,z}$ are the out-of-plane modal phase-matching coefficients. Equation~\eqref{eq:8} shows that the TH transmittance is composed of a helicity-dependent direct term ($\propto I_{1,z}$) that does not involve excitation of TH modes, and a helicity-independent resonant term ($\propto I_{2,z}$). 

Using the definition ${\rm CD}^{(3)}=(T^{(3)}_{+}-T^{(3)}_{-})/(T^{(3)}_{+}+T^{(3)}_{-})$ together with Eq.~\eqref{eq:8}, the nonlinear TH CD can be written in the compact form
\begin{equation}
\mathrm{CD}^{(3)}=
\frac{
A\sin(4\phi)+B\sin(8\phi)
}{
1+C\cos(4\phi)+D\cos(8\phi)
},
\label{eq:9}
\end{equation}
where $A$, $B$, $C$, and $D$ are real coefficients determined by the relative amplitudes and phases of the helicity-dependent and helicity-independent contributions in Eq.~\eqref{eq:8}. In the isotropic limit $\Delta\bar{\chi}^{(3)}=0$ the TH CD vanishes due to the vanishing coefficients $A$ and $B$, see Supporting Information, Section \ref{sm:subsec:nonlinear-cd}. The coefficients $B$ and $D$, which govern the eightfold angular harmonics in Eq.~\eqref{eq:9}, arise from the interference of the resonant response at the TH frequency and the direct nonlinear scattering pathway in Eq.~\eqref{eq:8}, whereas the direct nonlinear scattering pathway alone produces only fourfold angular harmonics. Therefore, the relative magnitudes of $B$ and $D$ compared with $A$ and $C$ provide a measure of the contribution of the TH resonances to the nonlinear CD.

We further analyse the performance of the designed realistic Si metasurface for TH CD generation numerically. Figure~\ref{fig:4}(a) shows the calculated $T^{(3)}_{\pm}$ for $\sigma=\pm 1$ pumping for different relative angles $\phi$ between the in-plane axes of the metasurface and those of $\hat{\chi}^{(3)}$ for the target metasurface design with $\delta=0.35$.
The signal is evaluated at the pump dipolar mode resonant frequency of $203$~THz and normalized to the respective TH transmission intensity of an unpatterned Si film of the same thickness. We can see that $T^{(3)}_{-}$ is strongly suppressed at $\phi=15^\circ$, which corresponds to the destructive interference of two helicity phase-matching pathways in Eq.~\eqref{eq:8}. Figure~\ref{fig:4}(b) shows the calculated $\mathrm{CD}^{(3)}$ and theoretical data fitted to Eq.~\eqref{eq:9} with $A=0.63$, $B=0.19$, $C=-0.41$, $D=0.17$, which are in good agreement. The maximal TH CD of $0.993$ is achieved at $\phi=15^\circ$. Figure~\ref{fig:4}(c) shows the frequency spectra of  $T^{(3)}_{\pm}$ for $\phi=15^\circ$ that confirm narrowband resonant generation of a helicity-selective harmonic signal. 

We next calculate the TH CD dependence on the hole size parameter $\delta$ and $\phi$. The frequency corresponds to the resonant frequency of the pump dipolar mode evaluated independently for each value of $\delta$, following the white dashed curve in Fig.~\ref{fig:3}(a). Figure~\ref{fig:4}(d) shows the comparison of calculated and fitted maps of $\mathrm{CD}^{(3)}$ with the fitting parameters shown in Fig.~\ref{fig:4}(e). The CD follows clear fourfold symmetry with respect to $\phi$ for $0.1\leq\delta\leq0.3$ and $0.4\leq\delta\leq0.45$, which is confirmed by small values of $B$ and $D$ in that range. For the range $0.3\leq\delta\leq0.4$, $B$ and $D$ are comparable with $A$ and $C$, thus, the eightfold dependence shapes the TH CD response. We note that in the range $0.3\leq\delta\leq0.4$ many TH modes cross the dispersion of the pump dipolar modes, as can be seen in Fig.~\ref{fig:3}(a). This confirms that the contribution of the TH modes is necessary for the eightfold dependence.

\begin{table}[htbp]
\centering
\caption{Material parameters and near-unity resonant $\mathrm{CD}^{(3)}$ for cubic nonlinear materials.}
\label{tab:1}
\renewcommand{\arraystretch}{1.4}
\setlength{\tabcolsep}{6pt}
\small
\begin{tabular}{lccccccccc}
\hline
Material
& $\omega/(2\pi)$ [THz]
& $n(\omega)$
& $n(3\omega)$
& $k(3\omega)$ 
& $3\chi^{(3)}_{1212}/\chi^{(3)}_{1111}$
& $\delta$
& $\varphi$ [$^\circ$] 
& $T^{(3)}_{\pm}$ enh.
& $\mathrm{CD}^{(3)}$ \\
\hline
Si           & $203$ & $3.50$ & $4.10$ & $0.04$ & $1.41$ & $0.35$ & $15$ & $3\times 10^3$ & $0.993$ \\
Ge           & $146$ & $4.10$ & $5.19$ & $0.53$ & $1.41 \eu^{\iu \pi /10}$ & $0.20$ & $20$ & $6\times 10^5$ & $0.998$ \\
C     & $232$ & $2.39$ & $2.42$ & $0$ & $1.30$ & $0.20$ & $23$ & $6\times 10^5$ & $0.998$ \\
Cu$_2$O      & $200$ & $2.63$ & $3.45$ & $0.12$ & $0.27$ & $0.20$ & $21$ & $5\times 10^5$ & $-0.994$ \\
CaF$_2$      & $202$ & $1.43$ & $1.44$ & $0$ & $1.54$ & $0.23$ & $18$ & $9\times 10^7$ & $-0.997$\\
\hline
\end{tabular}
\end{table}

We finally compare the performance of resonant nonlinear metasurfaces made of common cubic nonlinear materials across different near-infrared bands. Table~\ref{tab:1} shows the comparison of linear and nonlinear material properties, design parameters, and the resulting maximal value of normalised TH transmission and $\mathrm{CD}^{(3)}$ for Si, Ge, C (cubic diamond), Cu$_2$O and CaF$_2$ based on the same geometry of a perforated slab with a square lattice of circular air holes as in Fig.~\ref{fig:2}(b); see Supporting Information, Section \ref{sm:sec:materials} for more details and references. One can see that a near-unity resonant TH CD exceeding $99\%$ can be easily reached for materials with drastically different permittivity and nonlinear anisotropy parameters across a broad near-infrared range from $1300$~nm to $2050$~nm. This confirms that the proposed effect is universal to different materials and geometries, and establishes a simple CMOS-compatible design that provides robust generation of giant TH CD without breaking geometrical mirror symmetry or sophisticated design criteria. We note that different values of normalised TH transmission for each material are attributed to the mode $Q$ factor that can be controlled with $\delta$ and $n(\omega)$.


In conclusion, our results establish that near-unitary TH circular dichroism at the pump resonance does not require geometrically chiral structures or helicity-selective linear enhancement. Instead, for achiral nonlinear metasurfaces with cubic crystal symmetry, we have demonstrated that crystal orientation can be used directly to control the relative phase of several nonlinear generation pathways, allowing strong helicity selectivity to be achieved without complicating the metasurface geometry. We have shown that metasurfaces with rotational symmetry possess dipolar resonances supporting strong near-field components of both circular polarizations, enabling the required interference entirely through in-plane fields and removing the need for an additional out-of-plane nonlinear pathway commonly used in previous approaches for nonlinearity-induced chirality. This provides a particularly simple route to suppressing TH generation for one incident helicity while retaining resonantly enhanced emission for the other. The developed nonlinear temporal coupled-mode theory connects these pathways to the emitted harmonic signal and provides an analytical description of the crystal-orientation dependence, identifying the additional eightfold angular terms as signatures of contributions from TH resonances. The predicted resonant CD magnitudes exceeding \(99\%\) in several cubic materials with substantially different refractive indices and nonlinear anisotropies demonstrate that this design strategy is not specific to silicon. In silicon, it offers a simple, CMOS-compatible implementation that combines near-maximal helicity selectivity with resonant nonlinear enhancement. More broadly, these findings highlight how engineering material orientation and resonant near fields together can reduce the structural complexity required for nonlinear polarization control.

\subsection*{Supporting Information}
Linear temporal coupled mode theory derivation from Maxwell's equations for mode dispersions and resonant near fields; nonlinear polarization and helicity selection rules; nonlinear temporal coupled mode theory derivation; analytical derivation of third-harmonic transmission and near-unity third-harmonic circular dichroism; numerical results for Ge, diamond, Cu\(_2\)O, and CaF\(_2\).

\subsection*{Disclosures}
The authors declare that they have no competing financial interests or any other conflicts of interest.

\subsection*{Code, Data, and Materials Availability} 
The data used in this study are available from the corresponding author upon request.

\subsection*{Author Contributions}
K.K. conceived the idea. C.W., D.G., P.P. and K.K. performed the numerical simulations and analyzed the data. D.G., K.K. and L.W. supervised the project and provided theoretical guidance, contributed to the interpretation of the physical mechanisms, and shaped the overall research direction. C.W., D.G. and K.K. prepared the initial draft, and all authors contributed to discussions, revisions, and the final preparation of the manuscript.

\section*{Acknowledgements}
K.K. thanks Kristina Frizyuk for useful discussions. K.K. and P.P. acknowledge support from the Australian Research Council through a Discovery Early Career Researcher Award (Grant No. DE250100419). C.W., D.G., and L.W. acknowledge support from the National Research Foundation (NRF), Singapore (Grant Nos. NRF-CRP26-2021-0004 and NRF-CRP31-0007); the Ministry of Education (MOE), Singapore (Grant Nos. MOE-T2EP50223-0001, MOE-MOET32024-0005, and MOE-T2EP50125-0018); the Agency for Science, Technology and Research (A*STAR), Singapore (Grant No. MTC IRG M24N7c0083); and the Singapore University of Technology and Design (SUTD) under the Kickstarter Initiative (Grant No. SKI 2021-04-12).


\clearpage

\begin{figure}[htbp]
    \centering
    \includegraphics[width=5.5cm]{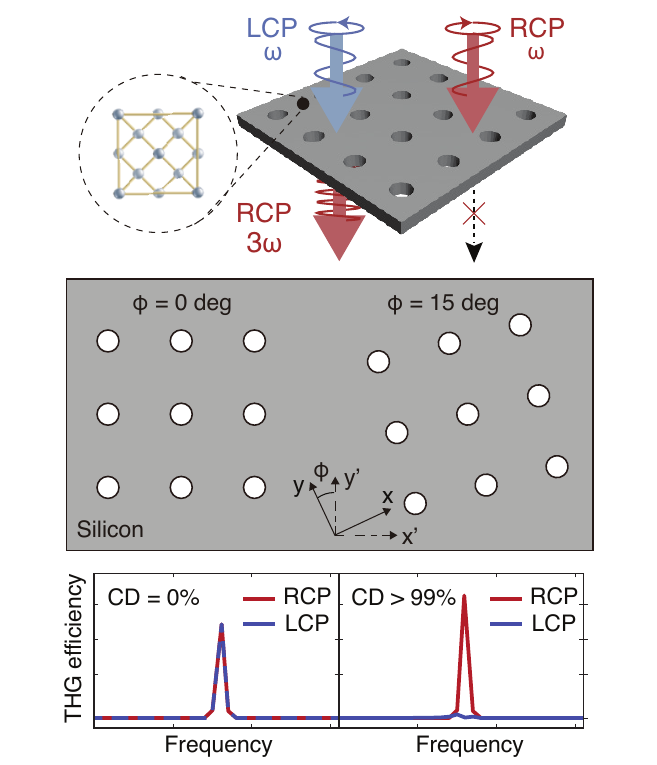}
    \caption{TOC Graphic}
\end{figure}

\StartSupportingInformation
\thispagestyle{empty}
\begingroup
\centering
{\fontsize{12}{14}\selectfont\bfseries
Supporting Information:\\
Crystal orientation enables near-unity nonlinear circular dichroism in resonant metasurfaces\par}
\vspace{12pt}
{\normalsize
Chi Wang\textsuperscript{1,*}, Dmitrii Gromyko\textsuperscript{1,*},
Polina Pantiukhina\textsuperscript{2}, Lin Wu\textsuperscript{1,$\dagger$},
and Kirill Koshelev\textsuperscript{2,$\ddagger$}\par}
\vspace{4pt}
{\small\itshape
\textsuperscript{1}Science, Mathematics, and Technology (SMT), Singapore University\\
of Technology and Design (SUTD), 8 Somapah Road, Singapore 487372\\
\textsuperscript{2}Department of Electronic Materials Engineering, Research School of Physics,\\
Australian National University, Canberra, ACT 2601, Australia\par}
\endgroup
\footnotetext[1]{These authors contributed equally.}
\footnotetext[2]{\href{mailto:lin_wu@sutd.edu.sg}{lin\_wu@sutd.edu.sg}}
\footnotetext[3]{\href{mailto:kirill.koshelev@anu.edu.au}{kirill.koshelev@anu.edu.au}}
\vspace{23pt}

\begin{quote}\textbf{Abstract:} Section \ref{sm:sec:linear} derives the linear temporal coupled-mode theory for resonant metasurfaces from Maxwell’s equations using the Lippmann–Schwinger formalism. It introduces the scattering channels and background field, eliminates the zeroth diffraction order as a radiative reservoir, derives the resonant mode expansion and coupling matrices, establishes the applicability criterion for the reservoir elimination, and diagonalizes the resulting Hamiltonian in the multipolar basis. Section \ref{sm:sec:linear} also derives the input and output coupling coefficients, their reciprocity, energy conservation, and time-reversal relations, and the corresponding linear scattering matrix. Section \ref{sm:sec:dispersion} extends the coupled-mode model to finite in-plane wavevectors near the $\Gamma$ point and derives the effective Hamiltonians and mode dispersions along the $\Gamma X$ and $\Gamma M$ directions for both $\rm s$- and $\rm p$-polarized guided modes. Section \ref{sm:sec:nearfields} derives the near-field distributions and helicity components of the resonant dipolar modes. Section \ref{sm:sec:selection} establishes the third-order nonlinear polarization and helicity selection rules for materials with $C_{4v}$ symmetry. Section \ref{sm:sec:nonlinear} develops the nonlinear temporal coupled-mode theory at the third-harmonic frequency, derives the nonlinear direct-scattering and resonant mode-coupling coefficients, obtains the effective nonlinear scattering matrix, and derives the analytical expression for nonlinear circular dichroism and its maximal value. Section \ref{sm:sec:materials} provides data for nonlinear transmission and TH CD for Ge, C (diamond), Cu$_2$O, and CaF$_2$. 
\end{quote}

\tableofcontents

\section{Linear temporal coupled mode theory}\label{sm:sec:linear}
\setcounter{equation}{0}

In this section, we derive the temporal coupled-mode theory given in the main text from Maxwell's equations following an approach similar to three-dimensional spatial coupled-mode theory~\cite{sm:liang2011three}. We consider a planar dielectric metasurface with fourthfold rotational symmetry ($C_{4v}$ point group symmetry) excited with a circularly polarized plane wave at normal incidence. The structure is periodic in the $xy$-plane with the relative permittivity $\eps(\bfr)$ and confined within a waveguide of finite thickness $h=2a$ along the $z$-axis. The permittivity is considered non-dispersive and lossless with the profile defined as:
\begin{equation}
\eps(\bfr)=
\begin{cases}
\eps(x,y), & |z|\le a,\\
1, & |z|>a.
\end{cases}
\end{equation}
The unit cell of the periodic lattice exhibits $C_{4v}$ point group symmetry. Assuming a harmonic time dependence of the fields proportional to $\eu^{-\iu \omega t}$, the electric field $\bfe(\bfr;k)$ at normal incidence (the $\Gamma$-point of the Brillouin zone) is governed by the vector Helmholtz equation:
\begin{equation}
\nabla\times\nabla\times\bfe(\bfr;k) - k^2\eps(\bfr)\bfe(\bfr;k) = 0,
\label{sm:eq:maxwell_real}
\end{equation}
where $k=\omega/c$ denotes the vacuum wavenumber and $c$ is the speed of light in vacuum.

To exploit the in-plane periodicity, we apply Bloch's theorem. The electric field, permittivity, and del operator are expanded into Fourier series over the reciprocal lattice vectors $\bfg=[g_x\ g_y\ 0]^{\mathsf{T}}$:
\begin{align}
&\bfe(\bfr;k)=\frac{1}{\sqrt{S_{\rm u}}}\sumg\eu^{\iu \bfg\cdot\bfr}\bfe_{\bfg}(z;k),\\
&\eps(\bfr)=\sumg\eu^{\iu \bfg\cdot\bfr} \eps_\bfg(z),\\
&\nabla f(\bfr)= \sum_\bfg\eu^{\iu \bfg\cdot\bfr}\nabla^\bfg_zf_\bfg(z),
\label{sm:eq:reciprocal_expansions}
\end{align}
where $S_{\rm u}$ represents the unit cell area, and the modified gradient operator is $\nabla^\bfg_z=[\iu g_x\ \iu g_y\ \partial_z]^{\mathsf{T}}$. Substituting these expansions into Eq.~\eqref{sm:eq:maxwell_real} yields the coupled wave equations:
\begin{equation}
\sum_{\bfg^\prime}k^2\eps_{\bfg-\bfg^\prime}(z)\bfe_{\bfg^\prime}(z;k)-\nabla^\bfg_z\times\nabla^\bfg_z\times\bfe_{\bfg}(z;k)=0.
\label{sm:eq:maxwell_k}
\end{equation}

\subsection{Scattering channels and background field}\label{sm:subsec:channels}

The scattering process is driven by an external background field $\bfe^{(\rm bg)}(z;k)$. We assume that the background field is a circularly polarized plane wave. Then, $\bfe^{(\rm bg)}(z;k)$ can be expanded in the basis of unperturbed scattering channels:
\begin{equation}
\bfe^{(\rm bg)}(z;k)=\sum_\sigma \uvect{\sigma } \sum_{u\in\{\mathrm{t,b}\}}\left(s^{\rm (in)}_{\sigma,u}(k){E}^{\rm (in)}_{u}(z;k)+s^{\rm (out,bg)}_{\sigma,u}(k){E}^{\rm (out)}_{u}(z;k)\right).
\label{sm:eq:bgexp}
\end{equation}
Here, $s^{\rm (in)}_{\sigma,u}$ and $s^{\rm (out,bg)}_{\sigma,u}$ denote the amplitudes of the incoming and outgoing waves. The index $u$ labels the propagation direction ($u={\rm t}$ for incidence from the top, $z<-a$; $u={\rm b}$ for incidence from the bottom, $z>a$), and $\sigma =\pm 1$ denotes the helical polarization. The unit polarization vectors are defined as $\uvect{\sigma }=(\uvect{x}-\iu\sigma \uvect{y})/\sqrt{2}$.

The channel functions ${E}^{\rm (in/out)}_{u}(z;k)$ satisfy the vacuum Helmholtz equation $(\partial_z^2 + k^2){E} = 0$, subject to the incoming and outgoing boundary conditions. Their explicit form is:
\begin{equation}
\begin{aligned}
&{E}^{\rm (in)}_{\rm t}(z;k) =\frac{ \eu^{ \iu k(z+a)} }{\sqrt{2}} H_\theta(-z),\quad {E}^{\rm (out)}_{\rm t}(z;k) =\frac{ \eu^{ - \iu k(z+a)} }{\sqrt{2}} H_\theta(-z), \\
&{E}^{\rm (in)}_{ \rm b}(z;k) =\frac{ \eu^{ -\iu k(z-a)} }{\sqrt{2}} H_\theta(z),\quad {E}^{\rm (out)}_{ \rm b}(z;k) =\frac{ \eu^{  \iu k(z-a)} }{\sqrt{2}} H_\theta(z),
\end{aligned}
\label{sm:eq:channel_funcs}
\end{equation}
where $H_\theta(z)$ is the Heaviside step function. We enforce orthonormality via the bilinear flux map $[E_1|E_2]_z$:
\begin{equation}
-\iu[{E}^{\rm (in)}_{ u}(z)|{E}^{\rm (out)}_{ u^\prime}(z)]_z = \iu[{E}^{\rm (out)}_{ u}(z)|{E}^{\rm (in)}_{ u^\prime}(z)]_z = \delta_{u,u^\prime}.
\label{sm:eq:orthonormality}
\end{equation}
The bilinear map is defined as:
\begin{equation}
[E_1(z)|E_2(z)]_z = \frac{1}{k}\left.\left(E_1E'_2-E'_1E_2\right)\right|_{z=-a}^{z=a}=\frac{1}{k}\intz\left(E_1E''_2-E''_1E_2\right).
\label{sm:eq:bilinear_map}
\end{equation}
The background amplitudes in Eq.~\eqref{sm:eq:bgexp} are extracted via projection:
\begin{equation}
s^{\rm (in)}_{\sigma ,u}(k) = \iu [{E}^{\rm (out)}_{u}(z;k)|\uvect{-\sigma}\cdot\bfe^{(\rm bg)}(z;k)]_z, \quad
s^{\rm (out,bg)}_{\sigma ,u}(k) = -\iu [{E}^{\rm (in)}_{u}(z;k)|\uvect{-\sigma }\cdot\bfe^{(\rm bg)}(z;k)]_z.
\end{equation}

\subsection{Lippmann--Schwinger's formalism}\label{sm:subsec:lippmann}

We employ a perturbation approach by separating the permittivity into a homogeneous effective waveguide $\overline{\eps}(z)$ and a periodic perturbation:
\begin{equation}
\eps_{\bfg}(z) = \delta_{\bfg,\mathbf 0}\overline{\eps}(z)+ \Delta\eps_{\bfg}(z), \quad \overline{\eps}(z)=
\begin{cases}
\veps, & |z|\le a,\\
1, & |z|>a.
\end{cases},\quad \Delta\eps_{\bfg}(z)=(1-\delta_{\bfg,\mathbf{0}})
\begin{cases}
\eps_{\bfg}, & |z|\le a,\\
0, & |z|>a.
\end{cases}
\end{equation}
This ensures $\Delta\eps_{\mathbf{0}}(z)=0$. The Maxwell operator for the effective waveguide is $\opp{L}_{\bfg}(z;k) = k^2\overline{\eps}(z)\opp{I} - \nabla^\bfg_z\times\nabla^\bfg_z\times$. Equation~\eqref{sm:eq:maxwell_k} becomes:
\begin{equation}
\opp{L}_{\bfg}(z;k)\bfe_{\bfg}(z;k) = -k^2\sum_{\bfg^\prime}\Delta\eps_{\bfg-\bfg^{\prime}}(z)\bfe_{\bfg^\prime}(z;k).
\end{equation}
The formal solution is obtained via the Lippmann-Schwinger equation using the dyadic Green's function $\opp{G}_{\bfg}(z,z^\prime;k)$:
\begin{equation}
\bfe_{\bfg}(z;k) = \bfe^{\rm (inc)}_{\bfg}(z;k) - k \intzp \opp{G}_{\bfg}(z,z^\prime;k) \sum_{\bfg^\prime\ne \bfg} \Delta\eps_{\bfg-\bfg^{\prime}}(z^\prime)\bfe_{\bfg^\prime}(z^\prime;k).
\label{sm:eq:lippmann_schwinger}
\end{equation}
The homogeneous solution $\bfe^{\rm (inc)}_{\bfg}(z;k)$ will be identified from the background field below, see Eq.~\eqref{sm:eq:source_exp}.

The Green's function satisfies $\opp{L}_{\bfg}\opp{G}_{\bfg} = k\opp{I}\delta(z-z^\prime)$ and is related to the real space GF as:
\begin{equation}
\opp{G}(\bfr,\bfr^\prime;k)=\frac{1}{S_{\mathrm{u}}} \sum_{\bfg} \mathrm{e}^{\mathrm{i}\bfg\cdot(\bfr-\bfr^{\prime})} \opp{G}_{\bfg}\left(z, z^{\prime};k\right).     
\end{equation}
For the zeroth order ($\bfg=\mathbf{0}$), it takes the explicit form:
\begin{equation}
\opp{G}_{\mathbf{0}}(z,z^\prime;k) = \frac{\delta(z-z^{\prime})}{\sqrt{\veps} q}\uvect{z}\otimes\uvect{z} + g_{\mathbf{0}}(z,z^\prime;k)\opp{I}_{\parallel},
\label{sm:eq:G0_explicit}
\end{equation}
where $q(k) = \sqrt{\veps}k$ is the wavenumber in the waveguide, $\opp{I}_{\parallel}=\opp{I}-\uvect{z}\otimes\uvect{z}$, and scalar Green's function $g_{\mathbf{0}}$ describes the Fabry-Perot poles of the waveguide:
\begin{equation}
g_{\mathbf{0}}(z,z^\prime;k)= \frac{E_{\rm t}(z_>;k)E_{\rm b}(z_<;k)}{\iu t_{\rm wg}(k)}.
\end{equation}
Here, $z_>=\operatorname{max}(z,z')$, $z_<=\operatorname{min}(z,z')$, the internal field functions $E_{u}(z;k)$ account for Fabry-Perot interference within the effective waveguide:
\begin{equation}
\begin{aligned}
& E_{\rm t}(z;k)=\frac{t_{\rm wg}(k)(\eu^{\iu q(z-a)}+{r}\eu^{-\iu q(z-a)})}{\sqrt{2}(1+r)}=\frac{[1+r_{\rm wg}(k)](\eu^{\iu q(z+a)}+{r}\eu^{4\iu qa}\eu^{-\iu q(z+a)})}{\sqrt{2}(1+r\eu^{4\iu qa})},\\
& E_{\rm b}(z;k)=\frac{t_{\rm wg}(k)(\eu^{-\iu q(z+a)}+{r}\eu^{\iu q(z+a)})}{\sqrt{2}(1+r)}=\frac{[1+r_{\rm wg}(k)](\eu^{-\iu q(z-a)}+{r}\eu^{4\iu qa}\eu^{\iu q(z-a)})}{\sqrt{2}(1+r\eu^{4\iu qa})},
\end{aligned}
\end{equation}
where $\tilde{r}(k)=\mathrm{e}^{2\iu qa}{r}$ is the phase-modulated Fresnel coefficient, with
\begin{equation}
{r}=\frac{\sqrt{\veps}-1}{\sqrt{\veps}+1},
\label{sm:eq:2.12}
\end{equation}
and the linear transmission and reflection amplitudes $t_{\rm wg}(k)$ and $r_{\rm wg}(k)$ are:
\begin{equation}
\begin{aligned}
&t_{\rm wg}(k)=\frac{[1-r^2]\mathrm{e}^{2\iu qa}}{[1-\tilde{r}^2(k)]},\\
&r_{\rm wg}(k)=\frac{r[\mathrm{e}^{4\iu qa}-1]}{[1-\tilde{r}^2(k)]}.
\end{aligned}
\end{equation}

By applying vector Green's identities, the homogeneous solution $\bfe^{\rm (inc)}_{\bfg}(z;k)$ of Eq.~\eqref{sm:eq:lippmann_schwinger} can be restored from the background field via zeroth order ($\bfg=\mathbf{0}$) Green's function at the top and bottom interfaces:
\begin{align}
&\bfe^{\rm (inc)}_\mathbf{0}(z;k) = -[\opp{G}_{\mathbf{0}}(z,z^\prime;k)|\bfe^{(\rm bg)}(z^\prime;k)]_{z^\prime}, \label{sm:eq:source_def}\\
&\bfe^{\rm (inc)}_{\bfg\ne \mathbf{0}}(z;k) = 0.
\end{align}
Explicitly, the source term expands as:
\begin{equation}
\bfe^{\rm (inc)}_\bfg(z;k) = \delta_{\bfg,\mathbf{0}}\sum_{\sigma,u}\uvect{\sigma }s^{\rm (in)}_{\sigma ,u}(k)E_{u}(z;k),
\label{sm:eq:source_exp}
\end{equation}
where we used $$[{E}^{\rm (in)}_{\rm t}(z;k)|E_{\rm b}(z;k)]_{z}=[{E}^{\rm (in)}_{\rm b}(z;k)|E_{\rm t}(z;k)]_{z}=\iu t_{\rm wg}(k)$$ and 
$$[{E}^{\rm (out)}_{\rm t}(z;k)|E_{\rm b}(z;k)]_{z}=[{E}^{\rm (out)}_{\rm b}(z;k)|E_{\rm t}(z;k)]_{z}=0.$$ 
This gives the formulas for $E_{u}(z;k)$:
\begin{equation}
E_{u}(z;k)=-[g_{\mathbf{0}}(z,z^\prime;k)|{E}^{\rm (in)}_{u}(z';k)]_{z'}.
\end{equation}

\subsection{Derivation of temporal coupled mode theory}\label{sm:subsec:linear-derivation}

\subsubsection{Elimination of the radiative reservoir}
To derive the resonant equations, we eliminate the zeroth order $\bfe_{\mathbf{0}}(z;k)$ treating it as a non-resonant reservoir that couples the resonant system to the radiation continuum. The applicability of this approximation is limited by the quality factor of the corresponding Fabry-Perot poles, as shown below in Eq.~\eqref{sm:eq:criterion}. We first isolate $\bfe_{\mathbf{0}}$ from Eq.~\eqref{sm:eq:lippmann_schwinger}:
\begin{align}
&\bfe_{\mathbf{0}}(z;k)=\bfe^{\rm (inc)}(z;k)-k   \intzp g_{\mathbf{0}}(z,z^\prime;k)\sum_{\bfg}\Delta\eps_{-\bfg}(z^\prime)\opp{I}_{\parallel}\bfe_{\bfg}(z^\prime;k),\label{sm:eq:2.7089}  \\
&\bfe_{\bfg}(z;k)=-k   \intzp \opp{G}_{\bfg}(z,z^\prime;k)\sum_{\bfg^\prime\ne\bfg,\mathbf{0}}\Delta\eps_{\bfg-\bfg^\prime}(z^\prime)\bfe_{\bfg^\prime}(z^\prime;k)-
k   \intzp \opp{G}_{\bfg}(z,z^\prime;k)\Delta\eps_{\bfg}(z^\prime)\bfe_{\mathbf{0}}(z^\prime;k).\label{sm:eq:2.709}    
\end{align}
Substituting Eq.~\eqref{sm:eq:2.7089} into Eq.~\eqref{sm:eq:2.709} yields the reduced system:
\begin{multline}
\bfe_{\bfg}(z;k) = - k\Delta\eps_{\bfg}\intzp \opp{G}_{\bfg}(z,z^\prime;k)\bfe^{\rm (inc)}(z^{\prime};k)-\\
k \sum_{\bfg^{\prime}\ne\mathbf{0}} \intzp \opp{G}_{\bfg}(z,z^\prime;k) \left[ \Delta\eps_{\bfg-\bfg^{\prime}}\bfe_{\bfg^\prime}(z^\prime;k) - k \Delta\eps_{\bfg} \Delta\eps_{-\bfg^{\prime}} \intzpp g_{\mathbf{0}}(z^\prime,z^{\prime\prime};k)\opp{I}_{\parallel}\bfe_{\bfg^\prime}(z^{\prime\prime};k) \right] .
\label{sm:eq:reduced_eq}
\end{multline}

\subsubsection{Resonant mode expansion}
We assume the system is excited near a guided mode resonance $k \approx k_{g}$ with TE or TM polarization. The dispersion relation of the guided modes is determined by the transverse resonance condition:
\begin{equation}
2 aq_{z,g}(k_{g})=\pi n+\operatorname{arg}[r_{\alpha,g}(k_{g})],\quad n\in \mathbb{Z},
\label{sm:eq:2.1201}
\end{equation}
where the Fresnel coefficients for oblique incidence at $\alpha=\rm s,p$ polarizations are:
\begin{equation}
{r}_{{\rm s},g}(k)=\frac{q_{z,g}(k)+\iu \varkappa_{z,g}(k)}{q_{z,g}(k)-\iu \varkappa_{z,g}(k)},\quad  {r}_{{\rm p},g}(k)=\frac{\iu\veps\varkappa_{z,g}(k)+q_{z,g}(k)}{\iu\veps\varkappa_{z,g}(k)-q_{z,g}(k)},
\label{sm:eq:2.12_oblique}
\end{equation}
with longitudinal wavenumber $q_{z,g}(k)=\sqrt{q^2-g^2}$ and transverse decay constant $\varkappa_{z,g}(k)=\sqrt{g^2-k^2}$.

Using the Mittag-Leffler theorem, the Green's function is approximated by its dominant pole:
\begin{equation}
\opp{G}_{\bfg}(z,z^\prime;k) \approx \frac{\bfe_{\bfg}(z)\otimes\bfe^{*}_{\bfg}(z^\prime)}{(k-k_{g})}.
\label{sm:eq:GFexp}
\end{equation}
Here, $\bfe_{\bfg}(z)$ satisfies $\opp{L}_{\bfg}(z;k_{g})\bfe_{\bfg}(z)=0$. The mode profiles are:
\begin{equation}
\bfe_{\bfg}(z)=\uvect{\alpha,\bfg}^\parallel \theta^{\rm \parallel}_{\alpha,g}(z)+\uvect{z} \theta^{\rm \perp}_{\alpha,g}(z),
\label{sm:eq:Egen}
\end{equation}
with polarization vectors $\uvect{{\rm s},\bfg}^\parallel=[\uvect{z}\times\uvect{\bfg}]$ and $\uvect{{\rm p},\bfg}^\parallel=\uvect{\bfg}$. The explicit field components are normalized according to $2\int_{-\infty}^\infty {\rm d}z\, \overline{\eps}(z)|\bfe_{\bfg}(z)|^2=1$, yielding for $|z|\le a$:
\begin{align}
&\theta^\|_{{\rm s},g}(z)=\frac{\cos{\left[q_{z,g}(k_g) z-\frac{\pi n}{2}\right]}}{\sqrt{2\veps a+\frac{2g^2}{ k_g^2\varkappa_{z,g}(k_g)}}}, \quad \theta^{\perp}_{{\rm s},g}(z)=0,\\
&\theta^\|_{{\rm p},g}(z)=\frac{q_{z,g}(k_g)\sin{\left[q_{z,g}(k_g) z-\frac{\pi n}{2}\right]}}{\sqrt{\veps }k_g\sqrt{2\veps a +\frac{2\veps}{ \varkappa_{z,g}(k_g)}}}, \quad \theta^{\perp}_{{\rm p},g}(z)=\frac{\iu g\cos{\left[q_{z,g}(k_g) z-\frac{\pi n}{2}\right]}}{\sqrt{\veps }k_g\sqrt{2\veps a +\frac{2\veps }{ \varkappa_{z,g}(k_g)}}}.
\end{align}
Substituting the ansatz $\bfe_{\bfg}(z;k) \approx a_{\bfg}(k) \bfe_{\bfg}(z)$ and Green's function expansion Eq.~\eqref{sm:eq:GFexp} into Eq.~\eqref{sm:eq:reduced_eq} leads to the amplitude equation:
\begin{multline}
a_{\bfg}(k)(k-k_{g}) =  -k \Delta\eps_{\bfg} \intz \bfe^{*}_{\bfg}(z) \cdot \bfe^{\rm (inc)}(z;k)-\\ 
k \sum_{\bfg^{\prime}\ne\mathbf{0}} a_{\bfg^{\prime}}(k) \intz \bfe^{*}_{\bfg}(z) \cdot \left[ \Delta\eps_{\bfg-\bfg^{\prime}}\bfe_{\bfg^{\prime}}(z) - k \Delta\eps_{\bfg} \Delta\eps_{-\bfg^{\prime}} \intzp g_{\mathbf{0}}(z,z^\prime;k)\opp{I}_{\parallel}\bfe_{\bfg^{\prime}}(z^\prime) \right],
\label{sm:eq:loos}
\end{multline}
where $a_{\bfg}(k)$ represent complex classical canonical variables in the Heisenberg picture.

\subsubsection{Symmetry analysis}

The mirror symmetries impose $\eps_{\bfg}=\eps_{-\bfg}$. The polarization vectors satisfy:
\begin{equation}
\begin{aligned}
&[\uvect{\alpha,\bfg}^\parallel\cdot\uvect{\alpha,\bfg^{\prime}}^\parallel]=[\uvect{\bfg}\cdot\uvect{\bfg^{\prime}}],\\
&[\uvect{\sigma }\cdot\uvect{\alpha,\bfg}^\parallel]= \xi_{\alpha,\sigma}  [\uvect{\sigma }\cdot\uvect{\bfg}],
\end{aligned}
\label{sm:eq:props}
\end{equation}
where $\xi_{\rm s,\sigma}=-\iu\sigma$, $\xi_{\rm p,\sigma}= 1$. In the vicinity of the guided-mode resonant frequency $k\approx k_{g}+O(\Delta\eps_\bfg)$, the interacting system is limited to four degenerate diffraction orders. The amplitude equations for four resonant amplitudes become:
\begin{equation}
a_{\bfg}(k)\left(k-k_{g}\right)
=\sum_{\bfg^\prime \ne \mathbf{0}}a_{\bfg^{\prime}}(k)\left\{\kappa^\perp_{|\bfg-\bfg^{\prime}|}+[\uvect{\bfg}\cdot\uvect{\bfg^{\prime}}]\left(\kappa^\parallel_{|\bfg-\bfg^{\prime}|}-\iu\gamma\right)\right\}+
\sqrt{2}\iu \sum_{\sigma,u} [\uvect{\sigma }\cdot\uvect{\bfg}] \xi_{\alpha,\sigma}D_{u}s^{\rm (in)}_{\sigma ,u}(k).
\label{sm:eq:2.3103}
\end{equation}
The coupling coefficients are defined explicitly as:
\begin{align}
&\kappa^\parallel_{|\bfg-\bfg^{\prime}|}=-k_{g}\Delta\eps_{|\bfg-\bfg^{\prime}|}\intz \left|\theta^\parallel_{\alpha,g}(z)\right|^2,
\label{sm:eq:kappa_ref}\\
&\kappa^\perp_{|\bfg-\bfg^{\prime}|}=-k_{g}\Delta\eps_{|\bfg-\bfg^{\prime}|}\intz \left|\theta^\perp_{\alpha,g}(z)\right|^2,\\
&\gamma=-k^2_{g}\Delta\eps^2_{g}  \intz\intzp \left[\theta^\parallel_{\alpha,g}(z)\right]^*\operatorname{Im}[g_{\mathbf{0}}(z,z^\prime;k_{g})]\theta^\parallel_{\alpha,g}(z^{\prime}) ,\label{sm:eq:Gamma_def}\\
&D_{u} =\frac{\iu k_{g}\Delta\eps_g}{\sqrt{2}}\intz\  \left[\theta^\parallel_{\alpha,g}(z)\right]^*E_u(z;k_{g}).
\end{align}
The real part of scalar Green's function $g_{\mathbf{0}}$  in Eq.~\eqref{sm:eq:Gamma_def} is omitted as a smaller order term compared to other real valued terms in Eq.~\eqref{sm:eq:2.3103}. 


\subsubsection{Criteria of zeroth-order elimination} 
The validity of treating the zeroth-order mode $\bfe_{\mathbf{0}}(z;k)$ as a non-resonant reservoir can be deduced from Eq.~\eqref{sm:eq:Gamma_def}. If the guided-mode frequency $k_{g}$ lies in the vicinity of the frequency $k_0$ of a Fabry-Perot mode, the zeroth order scalar Green's function can be approximated by the corresponding pole using the Mittag-Leffler theorem: $g_{\mathbf{0}}(z,z^\prime;k_{g}) \approx { \uvect{\mathbf{0}}\theta_0(z)\otimes\uvect{\mathbf{0}}^*\theta_0(z^\prime)}/{(k_{g}-k_0+\iu \gamma_0)}$. Here, $\gamma_0$ is the inverse radiative lifetime of the mode, and $\uvect{\mathbf{0}}\theta_0(z)$ is the eigenvector of $\opp{L}_{\mathbf{0}}(z;k_{0}-\iu \gamma_0)$. Therefore, the applicability criterion reduces to 
\begin{equation}
k_{g}\left|\Delta\eps_{g}[\uvect{\alpha,\bfg}^\parallel\cdot\uvect{\mathbf{0}}]\int_{-a}^{a}{\rm d}z\,   \left[\theta^\parallel_{\alpha,g}(z)\right]^*\theta_{0}(z)\right|\ll \left|k_{g}-k_0+\iu \gamma_0\right|.  
\label{sm:eq:criterion}
\end{equation}
In the limit $k_{g}\approx k_0$, $\theta_{g}(z)\approx\theta_{0}(z)\approx 1/\sqrt{\veps a}$, this criterion reduces to $\Delta\eps_{g}\ll 1/Q_0$, where $Q_0=k_0/(2\gamma_0)$ is the Fabry-Perot mode quality factor.

\subsubsection{Angular ordering and TCMT formulation}
The diffracted orders can be angularly counter-clockwise ordered: $\uvect{\bfg_l}=\cos(\pi l/2)\uvect{x}+\sin(\pi l/2)\uvect{y}$, where $l=0,\ldots,3$. The dot product becomes $[\uvect{\bfg_l}\cdot\uvect{\bfg_{l^\prime}}]=\cos[\pi (l^\prime-l)/2]$. Equation~\eqref{sm:eq:2.3103} transforms into:
\begin{equation}
a_{\bfg_l}(k)\left(k-k_{g}\right)
=\sum_{l^\prime=0}^{3}a_{\bfg_{l^{\prime}}}(k)\left\{\kappa^\perp_{|\bfg_l-\bfg_{l^{\prime}}|}+\left(\kappa^\parallel_{|\bfg_l-\bfg_{l^{\prime}}|}-\iu\gamma\right)\cos\left[ \frac{\pi(l^\prime-l)}{2}\right]\right\}
+\iu \sum_{\sigma, d} \iu^{-l\sigma } \xi_{\alpha,\sigma}D_{u}s^{\rm (in)}_{\sigma ,u}(k).   
\label{sm:eq:2.3104}
\end{equation}
Using the identity $|\bfg_{l+l^\prime}-\bfg_l|=|\bfg_{l+4-l^\prime}-\bfg_l|$, the coupling coefficients can be labeled as 
\begin{equation}
\kappa^{\parallel,\perp}_{l^\prime}\equiv\kappa^{\parallel,\perp}_{|\bfg_{l+l^\prime}-\bfg_l|}=\kappa^{\parallel,\perp}_{|\bfg_{l+4-l^\prime}-\bfg_l|}.    
\end{equation} 
In these notations, the system in Eq.~\eqref{sm:eq:2.3104} maps to the form of the first equation of the standard TCMT:
\begin{equation}
-\iu ka_{\bfg_l}(k)=-\iu\sum_{l^\prime}\left(\omega_{l,l^\prime}-\iu{\gamma}_{l,l^\prime}\right)a_{\bfg_{l^\prime}}(k)+\sum_{\sigma, d}  D_{\sigma,u,l}s^{\rm (in)}_{\sigma ,u}(k).
\label{sm:eq:tcmt}
\end{equation}
Here, the Hamiltonian and coupling matrices are:
\begin{align}
&\opp{\omega}= \begin{pmatrix}
k_{g} & \kappa_{1}^{\perp} &  \kappa^\perp_{2}-\kappa^\parallel_{2}  & \kappa_{1}^{\perp}  \\
\kappa_{1}^{\perp}  & k_{g} & \kappa_{1}^{\perp}& \kappa^\perp_{2}-\kappa^\parallel_{2} \\
\kappa^\perp_{2}-\kappa^\parallel_{2}  &\kappa_{1}^{\perp}& k_{g} &\kappa_{1}^{\perp}  \\
\kappa_{1}^{\perp}&\kappa^\perp_{2}-\kappa^\parallel_{2}   &\kappa_{1}^{\perp}& k_{g}   
\end{pmatrix},\label{sm:eq:mat1}\\
&\opp{\gamma}=  \gamma\begin{pmatrix}
1 & 0& -1 &0 \\
0 & 1& 0 &-1 \\
-1 & 0& 1 &0 \\
0 & -1& 0 &1 \\
\end{pmatrix},\label{sm:eq:mat2}\\
&D_{\sigma,u,l}=  \iu^{-l\sigma }\xi_{\alpha,\sigma} D_{u}.
\end{align}

\subsubsection{Second TCMT equation}
The coupling of the resonant amplitudes to the radiation continuum is analyzed by projecting the zeroth-order electric field $\bfe_{\mathbf{0}}(z=\pm a;k)$ onto the basis of input and output channels:
\begin{equation}
\bfe_{\mathbf{0}}(z=\pm a;k)= \sum_{\sigma',u'}\uvect{\sigma' }\left(s^{\rm (in)}_{\sigma',u'}(k){E}^{\rm (in)}_{u'}(z;k)+s^{\rm (out)}_{\sigma',u'}(k){E}^{\rm (out)}_{u'}(z;k)\right).   
\end{equation}
Invoking the reciprocity of the Green's function, $g_{\mathbf{0}}(z^\prime,z;k)=g_{\mathbf{0}}(z,z^\prime;k)$ and the property $[{E}^{\rm (in)}_{\rm t}(z;k)|E_{\rm t}(z;k)]_{z}=[{E}^{\rm (in)}_{\rm b}(z;k)|E_{\rm b}(z;k)]_{z}=\iu r_{\rm wg}(k)$, the amplitudes of the outgoing waves are derived as $s^{\rm (out)}_{\sigma ,u}(k) = -\iu [{E}^{\rm (in)}_{u}(z;k)|\uvect{-\sigma }\cdot\bfe_{\mathbf{0}}(z;k)]_z$ applied to Eq.~\eqref{sm:eq:2.7089}:
\begin{equation}
s^{\rm (out)}_{\sigma',u'}(k)=\sum_{\sigma,u}C^{\sigma',\sigma }_{u',u}(k)s^{\rm (in)}_{\sigma,u}(k)+\iu k_{g}  \Delta\eps_{g} \intz  E_{u'}(z;k)\sum_{l=0}^{3}[\uvect{-\sigma'}\cdot\bfe_{\bfg_l}(z;k)],
\label{sm:eq:2nd}
\end{equation}
where the elements of the background scattering matrix $C^{\sigma',\sigma}_{u',u}(k)$ are defined as:
\begin{equation}
\begin{aligned}
&C^{\sigma',\sigma}_{\rm t,t}(k)=C^{\sigma',\sigma}_{\rm b,b}(k)=\delta_{\sigma,\sigma' }r_{\rm wg}(k),\\
&C^{\sigma',\sigma}_{\rm t,b}(k)=C^{\sigma',\sigma}_{\rm b,t}(k)=\delta_{\sigma,\sigma' }t_{\rm wg}(k).
\end{aligned}
\end{equation}
Substituting the ansatz $\bfe_{\bfg}(z;k) \approx a_{\bfg}(k) \bfe_{\bfg}(z)$ into Eq.~\eqref{sm:eq:2nd} yields the full TCMT system in the non-diagonal basis:
\begin{align}
&-\iu ka_{\bfg_l}(k)=-\iu\sum_{l^\prime}\left(\omega_{l,l^\prime}-\iu{\gamma}_{l,l^\prime}\right)
a_{\bfg_{l^\prime}}(k)+\sum_{\sigma,u}  D_{\sigma,u,l}s^{\rm (in)}_{\sigma ,u}(k)\label{sm:eq:finalle},\\
&s^{\rm (out)}_{\sigma',u'}(k)=\sum_{\sigma,u}C^{\sigma',\sigma}_{u',u}(k)s^{\rm (in)}_{\sigma,u}(k)+\sum_l  K_{\sigma',u',l} a_{\bfg_l}(k),
\end{align}
where the outcoupling coefficients $K_{\sigma,u,l}$ are:
\begin{equation}
K_{\sigma,u,l}=\xi_{\alpha,-\sigma} D_{u} \iu^{l\sigma }.
\label{sm:eq:recip}
\end{equation}
We note the reciprocity property $K_{\sigma,u,l}=D_{-\sigma,u,l}$.

\subsubsection{Diagonalization via Discrete Fourier Transform}
The Hamiltonian matrices $\opp{\omega},\opp{\gamma}$ possess a circulant structure, reflecting the cyclic symmetry of the system. They are diagonalized by the Discrete Fourier Transform (DFT) matrix $\opp{U}$, with elements defined as $U_{\mu,l} = \iu^{l\mu}/2$. The indices denote the multipole order $\mu=0,\pm 1,2$ and the lattice vector index $l$.

The amplitudes of the eigenmodes (multipoles), defined by $\tilde{a}_{\mu}(k)=\sum_{l}{U}_{\mu,l}{a}_{l}(k)$, satisfy the following equations (decoupled at the $\Gamma$ point):
\begin{equation}
-\iu k\tilde{a}_{\mu}(k) = -\iu \left(\tilde{\omega}_{\mu} - \iu \tilde{\gamma}_{\mu}\right)\tilde{a}_{\mu}(k) + \sum_{\sigma,u} \tilde{D}_{\sigma,u,\mu} s^{\rm (in)}_{\sigma,u}(k).
\end{equation}
For a circulant Hamiltonian, the eigenvalues $\tilde{\omega}_{\mu} -\iu \tilde{\gamma}_{\mu}=\sum_{l,l^\prime}{U}_{\mu,l}\left(\omega_{l,l^\prime}-\iu \gamma_{l,l^\prime}\right)U^*_{\mu,l^\prime}$ are easily obtained as the Discrete Fourier Transform of its first row. The expressions for the eigenvalues simplify to:
\begin{align}
&\tilde{\omega}_{\mu} = k_{g} + 2 \operatorname{Re}[\iu^{\mu}]\kappa_{1}^{\perp} +
(-1)^{\mu}\left(\kappa_{2}^{\perp}- {\kappa_{ 2}^{\|}}\right).\\
&\tilde{\gamma}_{\mu} =   2\gamma\delta_{\mu,\pm 1}.
\end{align}
The transformed coupling coefficients $\tilde{D}_{\sigma,u,\mu}=\sum_{l}{U}_{\mu,l}{D}_{\sigma,u,l}$ and outcoupling coefficients $\tilde{K}_{\sigma,u,\mu}=\sum_{l}{U}^*_{\mu,l}{K}_{\sigma,u,l}$ satisfy the following selection rules:
\begin{equation}
\begin{aligned}
&\tilde{D}_{\sigma,u,\mu}=  2\xi_{\alpha,\sigma}  D_{u} \delta_{\mu,\sigma},\\
&\tilde{K}_{\sigma,u,\mu}= 2\xi_{\alpha,-\sigma}   D_{u} \delta_{\mu,\sigma}.
\end{aligned}
\label{sm:eq:DD}
\end{equation}

\subsubsection{Properties of coupling coefficients}
\paragraph{Reciprocity.} As follows from Eq.~\eqref{sm:eq:DD}, the reciprocity property in the diagonal basis~\cite{sm:ruan2012temporal} is:
\begin{equation}
\tilde{K}_{\sigma,u,\mu}=\sum_{\mu'}(\opp{U}\opp{U}^\mathsf{T})_{\mu,\mu'}\tilde{D}_{-\sigma,u,\mu'},    
\end{equation}
where $(\opp{U}\opp{U}^\mathsf{T})_{\mu',\mu}=\delta_{\mu,-\mu'}$ for doubly degenerate modes and $\delta_{\mu,\mu'}$ for non-degenerate modes.

\paragraph{Energy conservation.} The coupling coefficients are connected with the radiative loss rate as~\cite{sm:fan2003temporal}
\begin{equation}
\begin{aligned}
&\sum_{\sigma,u} \left|\tilde{D}_{\sigma,u,\mu}\right|^2= 2\tilde{\gamma}_{\mu},\\
&\sum_{u} \left|{D}_{u}\right|^2= \gamma.
\end{aligned}
\label{sm:eq:TCMT_D}
\end{equation}
In particular, for a metasurface in a symmetric environment: $|{D}_{\rm t}|^2=|{D}_{\rm b}|^2=\gamma/2$.

The proof is based on the following property:
\begin{equation}
\iint\limits_{-a}^a {\rm d}z{\rm d}z'\theta^\parallel_{\alpha,g}(z)\left(\frac{1}{2} \sum_u E^*_u(z;k)E_u(z';k)+\operatorname{Im}\left[\frac{E_{\rm t}(z_>;k)E_{\rm b}(z_<;k)}{\iu t_{\rm wg}(k)}\right]\right)\theta^\parallel_{\alpha,g}(z^{\prime})= 0.
\end{equation}

\paragraph{Time-reversal symmetry.} The coupling coefficients are connected to the direct scattering matrix~\cite{sm:ruan2012temporal}
\begin{equation}
\begin{aligned}
&\sum_{\sigma,u} C^{\sigma',\sigma}_{u',u}(k)\tilde{K}_{\sigma,u,\mu}^*= -\tilde{D}_{\sigma',u',\mu},\\
&\sum_{\sigma,u} C^{\sigma',\sigma}_{u',u}(k)D_{u}^*= -{D}_{u'}.
\end{aligned}
\label{sm:eq:TCMT_CD}
\end{equation}

The proof is based on the following properties:
\begin{equation}
\begin{aligned}
&\intz \theta^\parallel_{\alpha,g}(z)\left[r_{\rm wg}(k)E_{\rm t}^*(z;k)+t_{\rm wg}(k)E_{\rm b}^*(z;k)-E_{\rm t}(z;k)\right]= 0,\\    
&\intz \theta^\parallel_{\alpha,g}(z)\left[r_{\rm wg}(k)E_{\rm b}^*(z;k)+t_{\rm wg}(k)E_{\rm t}^*(z;k)-E_{\rm b}(z;k)\right]= 0.
\end{aligned}
\end{equation}

\subsection{Scattering matrix}\label{sm:subsec:linear-scattering}

The TCMT equations in the diagonal basis reduce to:
\begin{align}
&\tilde{a}_{\mu}(k)=\frac{\iu  \sum_{\sigma,u} \tilde{D}_{\sigma,u,\mu}s^{\rm (in)}_{\sigma,u}(k)}{\left(k-\tilde{\omega}_{\mu}+\iu\tilde{\gamma}_{\mu}\right)},\\
&s^{\rm (out)}_{\sigma',u'}(k)=\sum_{\sigma,u}C^{\sigma',\sigma }_{u',u}(k)s^{\rm (in)}_{\sigma,u}(k)+ \sum_{\mu} \tilde{K}_{\sigma',u',\mu}
\tilde{a}_{\mu}(k).
\end{align}
The resulting scattering matrix (S-matrix) defined as $s^{\rm (out)}_{\sigma',u'}(k)=\sum_{\sigma,u}S^{\sigma',\sigma }_{u',u}(k)s^{\rm (in)}_{\sigma,u}(k)$ is:
\begin{equation}
S^{\sigma',\sigma }_{u',u}(k)=C^{\sigma',\sigma }_{u',u}(k)+\iu  \sum_\mu\frac{\tilde{K}_{\sigma',u',\mu}\tilde{D}_{\sigma,u,\mu}}{\left(k-\tilde{\omega}_{\mu}+\iu \tilde{\gamma}_{\mu}\right)}.
\end{equation}
\begin{sloppypar}Using the selection rules in Eq.~\eqref{sm:eq:DD} and $\xi_{\alpha,-\sigma}\xi_{\alpha,\sigma}=1$, the second term in the RHS can be simplified: $\tilde{K}_{\sigma',u,\mu'}\tilde{D}_{\sigma,u,\mu}=4\delta_{\sigma,\sigma'}\delta_{\mu,\sigma}D_{u'}D_{u}$. As a result, the S-matrix is diagonal with respect to the polarization index:
\end{sloppypar}
\begin{equation}
S^{\sigma',\sigma }_{u',u}(k)=C^{\sigma',\sigma }_{u',u}(k)+\frac{4\iu \delta_{\sigma,\sigma'} D_{u'}D_{u}}{\left(k-k_{g} -\kappa_{2}^{\perp}+{\kappa_{ 2}^{\|}}+2\iu\gamma\right)}.
\end{equation}

\section{Mode dispersion in the vicinity of \texorpdfstring{$\Gamma$}{Gamma} point}\label{sm:sec:dispersion}

The proposed method can be extended to account for the dependence on the in-plane wavevector $\mathbf{k}$ in the vicinity of $\Gamma$ point within the perturbation theory $|\mathbf{k}|\ll k_g$. Linear and quadratic corrections can be added to the RHS of Eq.~\eqref{sm:eq:loos} in the following form
\begin{equation}
\ldots+\frac{2a_\bfg(k)}{k}\int_{-\infty}^\infty {\rm d}z\, \left\{[\mathbf{k}\cdot \bfg]\left|\bfe_{\bfg}\right|^2 -[\mathbf{k}\cdot\bfe_{\bfg}] [\bfg\cdot\bfe_{\bfg}]+
[\mathbf{k}\cdot\bfe_{\bfg}]^2\frac{\iu {\rm d}}{2{\rm d} z}\left(\frac{[\uvect z \cdot\bfe_{\bfg}]}{[\mathbf{k}\cdot\bfe_{\bfg}]}\right)+\frac{1}{2}\left|\mathbf{k}\times\bfe_{\bfg}\right|^2\right\}.      
\end{equation}
   
Then, Eq.~\eqref{sm:eq:finalle} transforms to 
\begin{multline}
-\iu ka_{\bfg_l}(k)=-\iu\sum_{l^\prime}\left[{\omega}_{l,l^\prime}-\iu{\gamma}_{l,l^\prime}\right]a_{\bfg_{l^\prime}}(k)+\sum_{\sigma,u}  D_{\sigma,u,l}s^{\rm (in)}_{\sigma,d }(k)+\\
\left(2\sqrt{2}g\operatorname{Re}[k_+\iu^{-l}]X_\alpha +2k_+k_-Y_\alpha+2\operatorname{Re}[k^2_+\iu^{-2l}]Z_\alpha\right) a_{\bfg_{l}}(k),
\end{multline}
where the coupling matrices are:
\begin{equation}
\begin{aligned}
&X_{\rm s}=\frac{1}{k_g}\int_{-\infty}^\infty {\rm d}z\,   \left|\theta^\parallel_{\rm s,g}(z)\right|^2,\quad X_{\rm p}=\frac{1}{k_g}\int_{-\infty}^\infty {\rm d}z\,\left\{   \left|\theta^\perp_{\rm p,g}(z)\right|^2+\left|\theta^\parallel_{\rm p,g}(z)\right|^2\frac{\iu {\rm d}}{(2g){\rm d} z}\left(\frac{\theta^\perp_{\rm p,g}(z)}{\theta^\parallel_{\rm p,g}(z)}\right)\right\},\\ 
&Y_{\rm s}= \frac{1}{k_g}\int_{-\infty}^\infty {\rm d}z\,   \left|\theta^\parallel_{\rm s,g}(z)\right|^2,\quad Y_{\rm p}= \frac{1}{k_g}\int_{-\infty}^\infty {\rm d}z\,   \left\{\left|\theta^\parallel_{\rm p,g}(z)\right|^2+2\left|\theta^\perp_{\rm p,g}(z)\right|^2\right\},\\
&Z_{\rm s}= \frac{1}{k_g}\int_{-\infty}^\infty {\rm d}z\,   \left|\theta^\parallel_{\rm s,g}(z)\right|^2,\quad Z_{\rm p}= -\frac{1}{k_g}\int_{-\infty}^\infty {\rm d}z\,   \left|\theta^\parallel_{\rm p,g}(z)\right|^2.
\end{aligned}    
\end{equation}
Direct evaluation of $X_{\alpha}$
gives 
\begin{equation}
\begin{aligned}
&X_{\rm s}=\frac{\left[1+\frac{1}{a\varkappa_{z,g}}\right]}{2k_g\veps\left[1+\frac{g^2}{a\varkappa_{z,g}q_g^2}\right]},\\
&X_{\rm p}=\frac{\left[1+\frac{ g^2[q_g^2-\varkappa^2_{z,g}]}{a\varkappa_{z,g}k^2_g(\veps \varkappa^2_{z,g}+g^2)}\right]}{2k_g\veps\left[1+\frac{ g^2}{a\varkappa_{z,g}(\veps \varkappa^2_{z,g}+g^2)}\right]}.
\end{aligned}
\end{equation}

In the basis that diagonalizes the Hamiltonian at the $\Gamma$ point, the $\mathbf{k}$-dependent terms are transformed so the resulting TCMT equation reads:
\begin{multline}
-\iu k\tilde{a}_{\mu}(k) = -\iu \left[\tilde{\omega}_{\mu} - \iu \tilde{\gamma}_{\mu}+2Y_\alpha k_+k_-\right]\tilde{a}_{\mu}(k)+\sum_{\sigma,u} \tilde{D}_{\sigma,u,\mu} s^{\rm (in)}_{\sigma,u}(k)+\\
\sum_{\mu'}\left\{\sqrt{2}gk_+\left(\delta_{\mu,\mu'+1}+\delta_{\mu,\mu'-3}\right)+\sqrt{2}gk_-\left(\delta_{\mu,\mu'+3}+\delta_{\mu,\mu'-1}\right)\right\} X_\alpha\tilde{a}_{\mu'}(k)+\\
\sum_{\mu'}Z_\alpha\left[k^2_++k^2_-\right]\left(\delta_{\mu,\mu'+2}+\delta_{\mu,\mu'-2}\right) \tilde{a}_{\mu'}(k).
\end{multline}
The matrix form of the Hamiltonian $\tilde{H}_\mu(\mathbf{k})$ is
\begin{multline}
\tilde{H}=\begin{pmatrix}
k_{g} +2\kappa_1^\perp+\kappa_2^\perp-\kappa^\|_{2}& 0 &  0 & 0  \\
0 & k_{g} -\kappa_2^\perp+\kappa^\|_{2}-2\iu\gamma & 0 & 0\\
0 & 0 & k_{g} -\kappa_2^\perp+\kappa^\|_{2}-2\iu\gamma & 0  \\
0 & 0 & 0 & k_{g} -2\kappa_1^\perp+\kappa_2^\perp-\kappa^\|_{2}    
\end{pmatrix}\\
+\begin{pmatrix}
2Y_\alpha k_+k_- & \sqrt{2}gX_\alpha k_- &  \sqrt{2}gX_\alpha k_+  & Z_\alpha (k_+^2+k_-^2)  \\
\sqrt{2}gX_\alpha k_+  & 2Y_\alpha k_+k_- & Z_\alpha(k_+^2+k_-^2) & \sqrt{2}gX_\alpha k_- \\
\sqrt{2}gX_\alpha k_-  & Z_\alpha(k_+^2+k_-^2) & 2Y_\alpha k_+k_-  &\sqrt{2}gX_\alpha k_+  \\
Z_\alpha(k_+^2+k_-^2)&\sqrt{2}gX_\alpha k_+   &\sqrt{2}gX_\alpha k_- & 2Y_\alpha k_+k_-    
\end{pmatrix}.    
\label{sm:eq:hma}
\end{multline}

We can further transform the Hamiltonian along high-symmetry directions defined by the polar angle $\phi$ in
$k_\pm=|\bfk| e^{\pm \iu \phi}/\sqrt{2}$. The corresponding transformation matrix is
\begin{equation}
U_\phi=\left(\begin{array}{cccc}
\cos{\beta} & 0 & 0 & \sin{\beta} \\
0 & \frac{\eu^{\iu \phi}}{\sqrt{2}} & \frac{\eu^{\iu \phi}}{\sqrt{2}} & 0 \\
0 & \frac{\eu^{-\iu \phi}}{\sqrt{2}} & -\frac{\eu^{-\iu \phi}}{\sqrt{2}} & 0 \\
-\sin{\beta} & 0 & 0 & \cos{\beta}
\end{array}\right) .
\label{sm:eq:U1}
\end{equation}

Then, $\tilde{H}_\phi=U^\dagger_\phi \tilde{H}U_\phi=\tilde{H}^{\rm (TE)}_\phi\oplus \tilde{H}^{\rm (TM)}_\phi$. 

\paragraph{s-polarization.} In this case, $N=X_\alpha=Y_\alpha=Z_\alpha$ and $\kappa_1^\perp=\kappa_2^\perp=0$. For $\Gamma X$ direction ($\phi=0$) we use $\beta=\pi/4$ and get:
\begin{align}
&\tilde{H}^{\rm (TE)}_{\Gamma X}= \begin{pmatrix}
k_{g} -\kappa^\|_{2}+2N |\bfk|^2 & 2g N |\bfk| &   0  \\
2g N |\bfk| & k_{g} +\kappa^\|_{2}-2\iu\gamma +2N |\bfk|^2 & 0\\
0 & 0  & k_{g} -\kappa^\|_{2}    
\end{pmatrix}   ,\\
&\tilde{H}^{\rm (TM)}_{\Gamma X}=k_{g} +\kappa^\|_{2}-2\iu\gamma      .
\end{align}

For $\Gamma M$ direction ($\phi=45^\circ$), we use $\beta=0$ and get:
\begin{align}
&\tilde{H}^{\rm (TE)}_{\Gamma M}= \begin{pmatrix}
k_{g} -\kappa^\|_{2}+N |\bfk|^2& \sqrt{2}g N |\bfk|  \\
\sqrt{2}gN|\bfk| & k_{g} +\kappa^\|_{2}-2\iu\gamma +N |\bfk|^2     
\end{pmatrix}   ,\\
&\tilde{H}^{\rm (TM)}_{\Gamma M}=\begin{pmatrix}
k_{g} +\kappa^\|_{2}-2\iu\gamma +N |\bfk|^2 & -\iu\sqrt{2} g N |\bfk|  \\
\iu \sqrt{2}g N |\bfk| &  k_{g} -\kappa^\|_{2}  +N |\bfk|^2 
\end{pmatrix} .
\end{align}

The corresponding eigenvalue dispersions in the perturbation limit are
\begin{equation}
\begin{aligned}
&\tilde{k}_0(\bfk)=k_0 -2\nu_g k_x^2-2\iu \gamma \nu_g[\kappa^\|_2]^{-1}k_x^2,\\
&\tilde{k}_{1,x}(\bfk)=k_1 +2\nu_g k_x^2 -2\iu \gamma\left(1-\nu_g[\kappa^\|_2]^{-1}k_x^2\right),\\
&\tilde{k}_{1,y}(\bfk)=k_1 +2\nu_g k_y^2 -2\iu \gamma \left(1-\nu_g[\kappa^\|_2]^{-1}k_y^2\right),\\
&\tilde{k}_{2}(\bfk)=k_0 -2\nu_g k_y^2-2\iu \gamma \nu_g[\kappa^\|_2]^{-1}k_y^2,
\end{aligned}    
\end{equation}
where $\nu_g = g^2N^2/\kappa^\|_{2}$.

\paragraph{p-polarization.} For $\Gamma X$ direction ($\phi=0$) we use $\beta=0$ and get:
\begin{align}
&\tilde{H}^{\rm (TE)}_{\Gamma X}= \begin{pmatrix}
k_{g} +2\kappa_1^\perp+\kappa_2^\perp-\kappa^\|_{2}& \sqrt{2}g X_\alpha |\bfk| &   Z_\alpha |\bfk|^2  \\
\sqrt{2}g X_\alpha |\bfk| & k_{g} -\kappa_2^\perp+\kappa^\|_{2}-2\iu\gamma +(Y_\alpha+Z_\alpha) |\bfk|^2 & \sqrt{2}g X_\alpha |\bfk|\\
Z_\alpha |\bfk|^2 & \sqrt{2}g X_\alpha |\bfk|  & k_{g} -2\kappa_1^\perp+\kappa_2^\perp-\kappa^\|_{2}    
\end{pmatrix}   ,\\
&\tilde{H}^{\rm (TM)}_{\Gamma X}=\begin{pmatrix}k_{g} -\kappa_2^\perp+\kappa^\|_{2}-2\iu\gamma +(Y_\alpha-Z_\alpha) |\bfk|^2 \end{pmatrix} .
\end{align}
The corresponding eigenvalue dispersions in the perturbation limit are
\begin{equation}
\begin{aligned}
&\tilde{k}_{0}= k_{g} +2\kappa_1^\perp+\kappa_2^\perp-\kappa^\|_{2} + \left(Y_\alpha+\frac{g^2X^2_\alpha}{(\kappa_1^\perp+\kappa_2^\perp-\kappa^\|_{2}+\iu \gamma)}\right)|\bfk|^2,\\    
&\tilde{k}_{1,x}= k_{g} -\kappa_2^\perp+\kappa^\|_{2}-2\iu\gamma +\left(Y_\alpha+Z_\alpha-\frac{2g^2X^2_\alpha (\kappa_2^\perp-\kappa^\|_{2}+\iu \gamma)}{\left(\kappa_2^\perp-\kappa^\|_{2}+\iu \gamma\right)^2-\left(\kappa_1^\perp\right)^2}\right) |\bfk|^2,   \\
&\tilde{k}_{1,y}= k_{g} -\kappa_2^\perp+\kappa^\|_{2}-2\iu\gamma +\left(Y_\alpha-Z_\alpha\right) |\bfk|^2 ,\\
&\tilde{k}_{2}= k_{g} -2\kappa_1^\perp+\kappa_2^\perp-\kappa^\|_{2} + \left(Y_\alpha+\frac{g^2X^2_\alpha}{(-\kappa_1^\perp+\kappa_2^\perp-\kappa^\|_{2}+\iu \gamma)}\right)|\bfk|^2.
\end{aligned}
\end{equation}

For $\Gamma M$ direction ($\phi=45^\circ$) we use $\beta=0$ and get:
\begin{align}
&\tilde{H}^{\rm (TE)}_{\Gamma M}= \begin{pmatrix}
k_{g} +2\kappa_1^\perp+\kappa_2^\perp-\kappa^\|_{2}+Y_\alpha |\bfk|^2& \sqrt{2}g X_\alpha |\bfk|  \\
\sqrt{2}g X_\alpha |\bfk| & k_{g} -\kappa_2^\perp+\kappa^\|_{2}-2\iu\gamma +Y_\alpha |\bfk|^2     
\end{pmatrix}   ,\\
&\tilde{H}^{\rm (TM)}_{\Gamma M}=\begin{pmatrix}
k_{g} -\kappa_2^\perp+\kappa^\|_{2}-2\iu\gamma +Y_\alpha |\bfk|^2 & -\iu\sqrt{2} g X_\alpha |\bfk|  \\
\iu \sqrt{2}g X_\alpha |\bfk| &  k_{g} -2\kappa_1^\perp+\kappa_2^\perp-\kappa^\|_{2}+Y_\alpha |\bfk|^2   
\end{pmatrix} .
\end{align}
The corresponding eigenvalue dispersions in the perturbation limit are
\begin{equation}
\begin{aligned}
&\tilde{k}_{0}= k_{g} +2\kappa_1^\perp+\kappa_2^\perp-\kappa^\|_{2} + \left(Y_\alpha+\frac{g^2X^2_\alpha}{(\kappa_1^\perp+\kappa_2^\perp-\kappa^\|_{2}+\iu \gamma)}\right)|\bfk|^2,\\    
&\tilde{k}_{1,x}= k_{g} -\kappa_2^\perp+\kappa^\|_{2}-2\iu\gamma +\left(Y_\alpha-\frac{g^2X^2_\alpha}{(\kappa_1^\perp+\kappa_2^\perp-\kappa^\|_{2}+\iu \gamma)}\right)|\bfk|^2,   \\
&\tilde{k}_{1,y}= k_{g} -\kappa_2^\perp+\kappa^\|_{2}-2\iu\gamma + \left(Y_\alpha-\frac{g^2X^2_\alpha}{(-\kappa_1^\perp+\kappa_2^\perp-\kappa^\|_{2}+\iu \gamma)}\right)|\bfk|^2,\\
&\tilde{k}_{2}= k_{g} -2\kappa_1^\perp+\kappa_2^\perp-\kappa^\|_{2} + \left(Y_\alpha+\frac{g^2X^2_\alpha}{(-\kappa_1^\perp+\kappa_2^\perp-\kappa^\|_{2}+\iu \gamma)}\right)|\bfk|^2,\\
\end{aligned}
\end{equation}


\section{Near-field of resonant dipolar modes}\label{sm:sec:nearfields}
The full field can be expanded in the new basis using the ansatz  
\begin{equation}
\bfe(\bfr;k)\approx \bfe^{\rm (inc)}(z;k)+\sum_\mu\tilde{a}_{\mu}(k) \bfe_{\mu}(\bfr), 
\label{sm:eq:anz}
\end{equation} 
where $\bfe_{\mu}(\bfr)=\bfe_{\mu,\mathbf{0}}(z)+\sum_{l=0}^{3}  \eu^{\iu \bfg_l\cdot\bfr}\bfe_{\mu,\bfg_l}(z)$ are the eigenmode fields with the Fourier components given by
\begin{align}
& \bfe_{\mu,\mathbf{0}}(z) =-k_{g}\Delta\eps_{g}   \intzp  g_{\mathbf{0}}(z,z^\prime;k)\sum_l U^*_{\mu,l}\opp{I}_{\parallel}\bfe_{\bfg_l}(z^\prime),\\
&\bfe_{\mu,\bfg_l}(z)= U^*_{\mu,l}\bfe_{\bfg_l}(z).   
\end{align} 

Under $\sigma$-polarized excitation, the excited field is determined solely by the dipolar modes, as dictated by the selection rules $\tilde{a}_{\mu}\propto \delta_{\mu,\sigma}$. The analysis of the zeroth diffraction order of dipolar modes requires the evaluation of the following sums:
\begin{equation}
\sum_{l=0}^{3}U^*_{\sigma,l}\uvect{\alpha,\bfg_l}^\parallel=\sqrt{2}\xi_{\alpha,-\sigma}\uvect{\sigma},
\end{equation}
where we employ the identities:
\begin{align}
&\Sigma_{1}= \frac{1}{4}\sum_{l=0}^{3}\left(\iu^{-l(\sigma-1)}+\iu^{-l(\sigma+1)}\right)=1,\\
&\Sigma_{2}= \frac{1}{4\iu}\sum_{l=0}^{3}\left(\iu^{-l(\sigma-1)}-\iu^{-l(\sigma+1)}\right)=-\iu\sigma.
\end{align}
Consequently, the zeroth diffraction order field associated with the dipolar modes exhibits circular polarization:
\begin{equation}
E^{\sigma_1}_{\mu=\sigma,\mathbf{0}}(z)  =\delta_{\sigma,\sigma_1}\sqrt{2} \xi_{\alpha,-\sigma}k_{g}\Delta\eps_g     \intzp g_{\mathbf{0}}(z,z^\prime;k_{g})\theta^{\rm \parallel}_{\alpha,g}(z^\prime),
\end{equation}
where $E^{\sigma_1}_{\mu=\sigma,\bfg}(z)=[\uvect{-\sigma_1}\cdot\bfe_{\mu=\sigma,\bfg}(z) ]$ is the co- or cross-polarized helical component of a dipole mode excited with a $\sigma$-polarized pump. The corresponding co- and cross-polarized in-plane components of higher diffraction orders for the dipolar modes are:
\begin{equation}
E^{\sigma_1}_{\mu=\sigma,\bfg_l}(z)  =\frac{ (\sigma\sigma_1)^{l}\xi_{\alpha,-\sigma_1}   \theta^{\rm \parallel}_{\alpha,g}(z)}{2\sqrt{2}}.
\label{sm:eq:modeEE}
\end{equation}
This expression indicates a mixing of helicity in the near-field region.

The full mode electric field in real space can be evaluated by direct summation of the corresponding diffraction orders:
\begin{equation}
E^{\sigma_1}_{\mu=\sigma}(\bfr) = \frac{\xi_{\alpha,-\sigma_1}}{\sqrt{2}} \left\{\theta^{\rm \parallel}_{\alpha,g}(z)\left[\cos{(gx)}+(\sigma\sigma_1)\cos{(gy)}\right]+2\delta_{\sigma,\sigma_1}k_{g}\Delta\eps_g    \int_{-a}^a {\rm d}z^\prime\, g_{\mathbf{0}}(z,z^\prime;k_{g})\theta^{\rm \parallel}_{\alpha,g}(z^\prime)\right\}
\label{sm:eq:sincos}
\end{equation}

\section{Nonlinear polarization selection rules}\label{sm:sec:selection}
The nonlinear polarization is determined by the third-order susceptibility tensor $\hat{\chi}^{(3)}(\bfr)$ and the full field at the pump frequency $k$:
\begin{equation}
\mathbf{P}_{\bfg}(z;3k)=\sum_{\bfg_1,\bfg_2,\bfg_3}\hat{\chi}^{(3)}_{\bfg_1+\bfg_2+\bfg_3-\bfg}(z)\vdots\, \bfe_{\bfg_1}(z;k)\otimes\bfe_{\bfg_2}(z;k)\otimes\bfe_{\bfg_3}(z;k),
\label{sm:eq:4.181}
\end{equation}
where we used $\hat{\chi}^{(3)}_{-\bfg}(z)=\hat{\chi}^{(3)}_{\bfg}(z)$ due to the mirror symmetries of $C_{4v}$ point group symmetry. We also note that the ratio of each Fourier component of the nonlinear susceptibility tensor to its average value is the same as that for the linear permittivity profile:
\begin{equation}
\hat{\chi}^{(3)}_{\bfg}(z)= \hat{\chi}^{(3)}_{\mathbf{0}}(z)\left[\delta_{\bfg,\mathbf{0}}+\Delta\eps_{\bfg}(z)/\veps\right]. 
\label{sm:eq:epss}
\end{equation}

We can project Eq.~\eqref{sm:eq:4.181} onto the circular basis as ${P}^{\sigma_4}_{\bfg}(z;3k)=[\uvect{-\sigma_4}\cdot\mathbf{P}_{\bfg}(z;3k)]$, then
\begin{equation}
{P}^{\sigma_4}_{\bfg}(z;3k)=\sum_{\bfg_1,\bfg_2,\bfg_3}\sum_{\sigma_1,\sigma_2,\sigma_3}{\chi}^{(3)}_{\bfg_1+\bfg_2+\bfg_3-\bfg,{\sigma_4,\sigma_1,\sigma_2,\sigma_3}}(z) E^{\sigma_1}_{\bfg_1}(z;k) E^{\sigma_2}_{\bfg_2}(z;k) E^{\sigma_3}_{\bfg_3}(z;k),
\label{sm:eq:4.1811}
\end{equation}
where ${\chi}^{(3)}_{\bfg,{\sigma_4,\sigma_1,\sigma_2,\sigma_3}}(z)=\uvect{-\sigma_4}\cdot\hat{\chi}^{(3)}_{\bfg}\vdots\,\uvect{\sigma_1}\otimes\uvect{\sigma_2}\otimes  \uvect{\sigma_3}$.

The in-plane Cartesian coefficients of $\hat{\chi}^{(3)}$ for a material with the point group symmetry that includes $C_{4v}$ (e.g., crystalline silicon) are:
\begin{align}
&\chi^{(3)}_{xxxx}=\chi^{(3)}_{yyyy}=\chi^{(3)}_{1111},\\
&\chi^{(3)}_{xxyy}=\chi^{(3)}_{xyyx}=\chi^{(3)}_{yyxx}=\chi^{(3)}_{xyxy}=\chi^{(3)}_{yxyx}=\chi^{(3)}_{yxxy}=\chi^{(3)}_{1212}.
\end{align}
Direct evaluation in the helical basis rotated by an angle $\phi$ with respect to the Cartesian axes gives:
\begin{equation}
\chi_{\sigma_4,\sigma_1,\sigma_2,\sigma_3}^{(3)}=\eu^{\iu (\sigma_1+\sigma_2+\sigma_3-\sigma_4)\phi}\left[\chi^{(3)}_{1111}\frac{\left(1-\sigma_4\sigma_1\sigma_2\sigma_3\right)}{4}+\chi^{(3)}_{1212}\frac{\left(\sigma_4\left[\sigma_1+\sigma_2+\sigma_3\right]-\left[\sigma_1\sigma_2+\sigma_1\sigma_3+\sigma_2\sigma_3\right]\right)}{4}\right].
\label{sm:eq:4.222}
\end{equation}
Equation~\eqref{sm:eq:4.222} establishes the near-field selection rules for the helicity conversion, also know as total angular momentum projection conservation laws~\cite{sm:menshikov2026tailoring} and Neumann's principle~\cite{sm:konishi2014polarization}:
\begin{equation}
\sigma_4-(\sigma_1+\sigma_2+\sigma_3)=0,\pm 4,\ldots.
\end{equation}
The allowed nonlinear transitions include $\sigma_1=\sigma_2=\sigma_3=-\sigma_4$ and $\sigma_1=-\sigma_2=-\sigma_3=-\sigma_4$ plus all permutations:
\begin{equation}
\begin{aligned}
&\chi_{-\sigma_1,\sigma_1,\sigma_1,\sigma_1}^{(3)}=\eu^{4\iu\sigma_1\phi}\Delta\chi^{(3)},\\
&\chi_{-\sigma_1,\sigma_1,-\sigma_1,-\sigma_1}^{(3)}=\chi_{-\sigma_1,-\sigma_1,\sigma_1,-\sigma_1}^{(3)}=\chi_{-\sigma_1,-\sigma_1,-\sigma_1,\sigma_1}^{(3)}=\chi^{(3)}_{0},
\end{aligned}    
\end{equation}
where $\Delta\chi^{(3)}=(\chi^{(3)}_{1111}-3\chi^{(3)}_{1212})/2$, $\chi^{(3)}_{0}=(\chi^{(3)}_{1111}+\chi^{(3)}_{1212})/2$. Thus, the nonlinear polarization components are
\begin{equation}
{P}^{-\sigma_1}_{\bfg}(z;3k)=\sum_{\bfg_1+\bfg_2+\bfg_3=\bfg}\left[
\eu^{4\iu\sigma_1\phi}\Delta\chi^{(3)}
E^{\sigma_1}_{\bfg_1}(z;k)E^{\sigma_1}_{\bfg_2}(z;k)E^{\sigma_1}_{\bfg_3}(z;k)+3\chi^{(3)}_{0} E^{\sigma_1}_{\bfg_1}(z;k)E^{-\sigma_1}_{\bfg_2}(z;k)E^{-\sigma_1}_{\bfg_3}(z;k)
\right].
\label{sm:eq:4.181122}
\end{equation}

\section{Nonlinear temporal coupled mode theory}\label{sm:sec:nonlinear}
\subsection{Derivation of nonlinear temporal coupled mode theory}\label{sm:subsec:nonlinear-derivation}
\subsubsection{Elimination of the radiative reservoir}

For a metasurface made of a nonlinear material, Eqs.~(\ref{sm:eq:2.7089},\ref{sm:eq:2.709})  at the third-harmonic frequency can be written as
\begin{equation}
\begin{aligned}
&\bfe_{\mathbf{0}}(z;3k)=-3k   \intzp   g_{\mathbf{0}}(z,z^\prime;3k)\sum_{\bfg}\Delta\eps_{-\bfg}(z^\prime)\opp{I}_{\parallel}\bfe_{\bfg}(z^\prime;3k)-3k   \intzp g_{\mathbf{0}}(z,z^\prime;3k)\opp{I}_{\parallel}\mathbf{P}_{\mathbf{0}}(z';3k),\\
&\bfe_{\bfg}(z;3k)=-3k   \intzp \opp{G}_{\bfg}(z,z^\prime;3k)\sum_{\bfg^\prime\ne\bfg,\mathbf{0}}\Delta\eps_{\bfg-\bfg^\prime}(z^\prime)\bfe_{\bfg^\prime}(z^\prime;3k)-3k   \intzp \opp{G}_{\bfg}(z,z^\prime;3k)\Delta\eps_{\bfg}(z^\prime)\bfe_{\mathbf{0}}(z^\prime;3k)\\
&-3k   \intzp \opp{G}_{\bfg}(z,z^\prime;3k)\mathbf{P}_{\bfg}(z';3k).    
\end{aligned}
\label{sm:eq:3.1}
\end{equation}

\subsubsection{Resonant mode expansion}
The mode density at the harmonic frequency is normally high. We therefore assume that the TH frequency $3k\simeq 3k_g$ lies in the vicinity of some quasi-guided mode frequency $k^{(3)}_{g}-\iu \gamma^{(3)}_{\rm nr}$. Here, $k^{(3)}_{g}\simeq 3k_g$ and $\gamma^{(3)}_{\rm nr}$ describes mode absorption losses common at harmonic frequencies due to permittivity dispersion. For the sake of simplicity, we assume the mode in-plane wavevector is equal to $3\bfg$, where $|\bfg|=g$ as before.  Then, the excited field can be expanded as $\bfe_{3\bfg}(z;3k) \approx b_{3\bfg}(3k) \bfe^{(3)}_{3\bfg}(z)$ and the GF can be approximated as
\begin{equation}
\opp{G}_{3\bfg}(z,z^\prime;3k) \approx \frac{\bfe^{(3)}_{3\bfg}(z)\otimes\bfe^{(3),*}_{3\bfg}(z^\prime)}{(3k-k_{g}^{(3)}+\iu \gamma^{(3)}_{\rm nr})}.
\label{sm:eq:GFexpNL}
\end{equation}
Substituting these expressions into Eq.~\eqref{sm:eq:3.1} and performing the same transformations as in the linear regime, we arrive at the nonlinear TCMT system, where we assume that the pump polarization and direction are fixed to $\sigma$ and $u$, respectively,
\begin{equation}
\begin{aligned}
&-3\iu k b_{3\bfg_l}(3k)=-\iu\sum_{l^\prime}\left({\omega}^{(3)}_{l,l^\prime}-\iu{\gamma}^{(3)}_{l,l^\prime}-\iu{\gamma}^{(3)}_{\rm nr}\delta_{l,l^\prime}\right)
b_{3\bfg_{l^\prime}}(3k)+  D^{(3)}_{\sigma,u,l}[s^{\rm (in)}_{\sigma,u}(k)]^3,\\
&s^{\rm (out)}_{\sigma',u'}(3k)=C^{(3),\sigma',\sigma}_{u',u}(k)[s^{\rm (in)}_{\sigma,u}(k)]^3+\sum_l  K^{(3)}_{\sigma',u',l} b_{3\bfg_l}(3k).
\end{aligned}    
\label{sm:eq:tcmtnl}
\end{equation}
The nonlinear coupling matrices are
\begin{equation}
\begin{aligned}
&D^{(3)}_{\sigma,u,l}=3\iu k_{g}[s^{\rm (in)}_{\sigma,u}(k)]^{-3}
\sum_{\bar{\sigma}}\intz\left[E^{(3),\bar{\sigma}}_{3\bfg_l}(z)\right]^*P^{\bar{\sigma}}_{3\bfg_l}(z;3k_g),\\
&C^{(3),\sigma',\sigma}_{u',u}(k)=3\iu k[s^{\rm (in)}_{\sigma,u}(k)]^{-3}
\intz E_{u'}(z;3k)P^{\sigma'}_{\mathbf{0}}(z;3k_g),\\
&K^{(3)}_{\sigma,u,l}=\xi_{\alpha,-\sigma} D^{(3)}_{u} \iu^{l\sigma },\\
&D^{(3)}_{u}=\frac{3\iu k_{g}\Delta\eps_{3g}}{\sqrt{2}}\intz\  \left[\theta^{(3),\parallel}_{\alpha,3g}(z)\right]^*E_u(z;3k_{g}).
\end{aligned}
\end{equation}

The Hamiltonian matrices $\opp{\omega}^{(3)}$ and $\opp{{\gamma}}^{(3)}$ and corresponding; near-field coupling coefficients are given by Eqs.~(\ref{sm:eq:kappa_ref},\allowbreak\,\ref{sm:eq:Gamma_def},\allowbreak\,\ref{sm:eq:mat1},\allowbreak\,\ref{sm:eq:mat2}). Here, we assumed that the nonlinear polarization is generated by an $\rm s$-polarized mode at the pump frequency and thus it has no $z$-components contributing to the coupling coefficients.

\subsubsection{Diagonalization via Discrete Fourier Transform}
We next apply the DFT transformation with matrix $\opp{U}$, with elements defined as $U_{\mu,l} = \iu^{l\mu}/2$ to Eqs.~\eqref{sm:eq:tcmtnl}, and get
\begin{equation}
\begin{aligned}
&-3\iu k \tilde{b}_{\mu}(3k)=-\iu\left(\tilde{\omega}^{(3)}_{\mu}-\iu\tilde{\gamma}^{(3)}_{\mu}-\iu\tilde{\gamma}^{(3)}_{\rm nr}\right)
\tilde{b}_{\mu}(3k)+  \tilde{D}^{(3)}_{\sigma,u,\mu}[s^{\rm (in)}_{\sigma,u}(k)]^3\label{sm:eq:finalleNL},\\
&s^{\rm (out)}_{\sigma',u'}(3k)=C^{(3),\sigma',\sigma}_{u',u}(k)[s^{\rm (in)}_{\sigma,u}(k)]^3+\sum_\mu  \tilde{K}^{(3)}_{\sigma',u',\mu} \tilde{b}_{\mu}(3k),
\end{aligned}    
\end{equation}
where
\begin{equation}
\begin{aligned}
&\tilde{D}^{(3)}_{\sigma,u,\mu}=  3\iu k_{g}[s^{\rm (in)}_{\sigma,u}(k)]^{-3}
\sum_{\bar{\sigma}}\sum_l\intz\left[E^{(3),\bar{\sigma}}_{\mu,3\bfg_l}(z)\right]^*P^{\bar{\sigma}}_{3\bfg_l}(z;3k_g),\\
&\tilde{K}^{(3)}_{\sigma,u,\mu}= 2\delta_{\mu,\sigma}\xi_{\alpha,-\sigma}   D^{(3)}_{u} .
\end{aligned}
\label{sm:eq:DD12}
\end{equation}

\subsubsection{Nonlinear direct scattering coefficient}
Substituting Eqs.~(\ref{sm:eq:anz},\allowbreak\,\ref{sm:eq:4.181},\allowbreak\,\ref{sm:eq:4.181122}) into the expression for the nonlinear direct scattering matrix, we get
\begin{equation}
C^{(3),\sigma',\sigma }_{u',u}(k)=\left[\frac{\tilde{a}_{\sigma}(k)}{s^{\rm (in)}_{\sigma,u}(k)}\right]^3\sum_{\sigma_1,\sigma_2,\sigma_3} \chi^{(3)}_{\mathbf{0},\sigma',\sigma_1,\sigma_2,\sigma_3} J_{\sigma_1,\sigma_2,\sigma_3}.
\label{sm:eq:wows}
\end{equation}
Here, the nonlinear outcoupling (phase-matching) coefficient $J_{\sigma_1,\sigma_2,\sigma_3}(k)$ is 
\begin{equation}
J_{\sigma_1,\sigma_2,\sigma_3}= \sum_{\bfg_1,\bfg_2,\bfg_3}  3\iu k_g\left(\delta_{\bfg_1+\bfg_2+\bfg_3,\mathbf{0}}+\Delta\eps_{\bfg_1+\bfg_2+\bfg_3}\right)\intz E_{u'}(z;3k_g)E^{\sigma_1}_{\sigma,\bfg_1}(z)E^{\sigma_2}_{\sigma,\bfg_2}(z)E^{\sigma_3}_{\sigma,\bfg_3}(z).
\label{sm:eq:Ks}
\end{equation}
Assuming only combinations with $\bfg_1=\bfg_2=\bfg_3=\bfg_l$ contribute, we can write at the pump resonance $k\simeq k_\sigma$
\begin{equation}
C^{(3),\sigma',\sigma }_{u',u}(k)=\delta_{\sigma',-\sigma} \left(\frac{D_u}{\gamma}\right)^3I_{1,z,u'}\left[\eu^{4\iu\sigma\phi}\Delta\bar{\chi}^{(3)}+3\bar{\chi}_{0}^{(3)}\right],
\label{sm:eq:C1}
\end{equation}
where the out-of-plane phase-matching is defined as
\begin{equation}
I_{1,z,u'}=\frac{3\iu k_g \Delta\varepsilon_{3g}}{4\sqrt{2}}   \intz E_{u'}(z;3k_g)[\theta^{\rm \parallel}_{\alpha,g}(z)]^3.
\end{equation}

\subsubsection{Nonlinear mode coupling coefficient}
Using Eq.~\eqref{sm:eq:4.181122}, we can show that
\begin{equation}
\tilde{D}^{(3)}_{\sigma,u,\mu}=\left[\frac{\tilde{a}_\sigma(k_g)}{s^{\rm (in)}_{\sigma,u}(k_g)}\right]^{3}\sum_{\bar{\sigma}} \frac{1}{2}\left[\eu^{4\iu\bar{\sigma}\phi}\Delta\bar{\chi}^{(3)}I_{-\bar{\sigma},\bar{\sigma},\bar{\sigma},\bar{\sigma}}+3\bar{\chi}_{0}^{(3)}I_{-\bar{\sigma},\bar{\sigma},-\bar{\sigma},-\bar{\sigma}} \right], 
\label{sm:eq:nlDD21}
\end{equation}
where $I_{\sigma_4,\sigma_3,\sigma_2,\sigma_1}$ is defined as
\begin{equation}
I_{\sigma_4,\sigma_3,\sigma_2,\sigma_1}=6\iu k_g\intz\sum_l\left[E^{(3),\sigma_4}_{\mu,3\bfg_l}(z)\right]^*E^{\sigma_1}_{\sigma,\bfg_l}(z)E^{\sigma_2}_{\sigma,\bfg_l}(z)E^{\sigma_3}_{\sigma,\bfg_l}(z).
\label{sm:eq:II}
\end{equation}
Substituting Eq.~\eqref{sm:eq:modeEE} into Eq.~\eqref{sm:eq:II}, we can find 
\begin{equation}
I_{-\bar{\sigma},\bar{\sigma},\bar{\sigma},\bar{\sigma}}=I_{-\bar{\sigma},\bar{\sigma},-\bar{\sigma},-\bar{\sigma}}   =\frac{3\delta_{\mu,-\sigma}\iu k_g\xi_{\alpha',-\sigma}\xi^3_{\alpha,-\sigma}  }{8} \intz [\theta^{(3),\rm \parallel}_{\alpha',3g}(z)]^*[\theta^{\rm \parallel}_{\alpha,g}(z)]^3. 
\end{equation}
Therefore, only the cross-polarized modes with $\mu=-\sigma$ are excited at the harmonic frequency by the near-field nonlinear source. Thus, Eq.~\eqref{sm:eq:nlDD21} can be written as
\begin{equation}
\tilde{D}^{(3)}_{\mu}=\delta_{\mu,-\sigma}\left(\frac{ D_u}{\gamma}\right)^3I_{-1,1,1,1}\left[\cos{(4\phi)}\Delta\bar{\chi}^{(3)}+3\bar{\chi}_{0}^{(3)}\right].   
\label{sm:eq:C2}
\end{equation}

\subsection{Nonlinear scattering matrix}\label{sm:subsec:nonlinear-scattering}
The effective nonlinear S-matrix $S^{(3),\sigma',\sigma }_{u',u}(k)$ is defined by $s^{\rm (out)}_{\sigma',u'}(3k)=\sum_{\sigma,u}S^{(3),\sigma',\sigma }_{u',u}(k)[s^{\rm (in)}_{\sigma,u}(k)]^3$~\cite{sm:koshelev2024scattering}. From the nonlinear TCMT equations, it can be written as
\begin{equation}
S^{(3),\sigma',\sigma }_{u',u}(k)=  C^{(3),\sigma',\sigma}_{u',u}(k)+\sum_\mu\frac{\iu \tilde{K}^{(3)}_{\sigma',u',\mu}\tilde{D}^{(3)}_{\sigma,u,\mu}}{\left(3k-\tilde{\omega}^{(3)}_{\mu}+\iu\tilde{\gamma}^{(3)}_{\mu}+\iu\tilde{\gamma}^{(3)}_{\rm nr}\right)}.
\label{sm:eq:C3}
\end{equation} 
Substituting Eqs.~(\ref{sm:eq:C1},\allowbreak\,\ref{sm:eq:C2}) into Eq.~\eqref{sm:eq:C3}, we obtain
\begin{equation}
S^{(3),\sigma',\sigma }_{u',u}(k)=  \delta_{\sigma',-\sigma} \left(\frac{D_u}{\gamma}\right)^3\left\{I_{1,z,u'}\left[\eu^{4\iu\sigma\phi}\Delta\bar{\chi}^{(3)}+3\bar{\chi}_{0}^{(3)}\right]+\frac{I_{2,z,u'}\left[\cos{(4\phi)}\Delta\bar{\chi}^{(3)}+3\bar{\chi}_{0}^{(3)}\right]}{(3k-\tilde{\omega}^{(3)}_{-\sigma}+\iu\tilde{\gamma}^{(3)}_{-\sigma}+\iu\tilde{\gamma}^{(3)}_{\rm nr,-\sigma})}\right\},
\label{sm:eq:Cs3}
\end{equation}
where
\begin{equation}
I_{2,z,u'} =-\frac{3k_g}{4} D_{u'}^{(3)}\intz [\theta^{(3),\rm \parallel}_{\alpha',3g}(z)]^*[\theta^{\rm \parallel}_{\alpha,g}(z)]^3. 
\end{equation}

\subsection{Nonlinear circular dichroism}\label{sm:subsec:nonlinear-cd} 

The nonlinear circular dichroism is defined via the cross-polarized emission:
\begin{equation}
{\rm CD}_{u',u}^{(3)}= \frac{|S^{(3),-1,+1 }_{u',u}|^2-|S^{(3),+1,-1 }_{u',u}|^2}{|S^{(3),-1,+1 }_{u',u}|^2+|S^{(3),+1,-1}_{u',u}|^2}.
\end{equation}

For compactness, we introduce
\begin{align}
x&=4\phi,
\\
X&=\Delta\bar{\chi}^{(3)},
\\
Y&=\bar{\chi}_{0}^{(3)},
\\
\mathcal{I}_{u',u}&=\left(\frac{D_u}{\gamma}\right)^3I_{1,z,u'},
\\
\mathcal{R}_{u',u}
&=
\left(\frac{D_u}{\gamma}\right)^3\frac{
I_{2,z,u'}
}{
\left[3k-\tilde{\omega}^{(3)}_{1}
+\iu\left(
\tilde{\gamma}^{(3)}_{1}
+\tilde{\gamma}^{(3)}_{\rm nr,1}
\right)
\right]},
\\
L_{u',u}&=\mathcal{I}_{u',u}+\mathcal{R}_{u',u}.
\end{align}
Then,
\begin{equation}
|S^{(3),-\sigma,\sigma }_{u',u}|^2
=
\left|
L_{u',u}\left(3Y+X\cos x\right)
+
\iu\sigma\mathcal{I}_{u',u}X\sin x
\right|^{2}.
\end{equation}
Introducing
\begin{equation}
P_{u',u}=L_{u',u}\left(3Y+X\cos x\right),
\qquad
Q_{u',u}=\mathcal{I}_{u',u}X\sin x,
\end{equation}
we have
\begin{equation}
|S^{(3),-\sigma,\sigma }_{u',u}|^2
=
|P_{u',u}|^{2}
+
|Q_{u',u}|^{2}
+
2\sigma\operatorname{Im}\left(P_{u',u}Q_{u',u}^{*}\right).
\end{equation}
Consequently, the circular dichroism
takes the form
\begin{equation}
{\rm CD}_{u',u}^{(3)}
=
\frac{
A_{u',u}\sin(4\phi)+B_{u',u}\sin(8\phi)
}{
1+C_{u',u}\cos(4\phi)+D_{u',u}\cos(8\phi)
},
\end{equation}
with the real coefficients
\begin{align}
A_{u',u}
&=
\frac{
6\operatorname{Im}
\left(
L_{u',u}\mathcal{I}_{u',u}^{*}YX^{*}
\right)
}{
\mathcal{N}_{u',u}
},
\\
B_{u',u}
&=
\frac{
|X|^{2}
\operatorname{Im}
\left(
\mathcal{R}_{u',u}\mathcal{I}_{u',u}^{*}
\right)
}{
\mathcal{N}_{u',u}
},
\\
C_{u',u}
&=
\frac{
6|L_{u',u}|^{2}
\operatorname{Re}
\left(
YX^{*}
\right)
}{
\mathcal{N}_{u',u}
},
\\
D_{u',u}
&=
\frac{
|X|^{2}
\left(
|\mathcal{R}_{u',u}|^{2}+2\operatorname{Re}
\left(
\mathcal{R}_{u',u}\mathcal{I}^*
\right)
\right)
}{
2\mathcal{N}_{u',u}
},
\end{align}
where 
\begin{equation}
\mathcal{N}_{u',u}
=
9|L_{u',u}|^{2}|Y|^{2}
+
\frac{|X|^{2}}{2}
\left(
|L_{u',u}|^{2}+|\mathcal{I}_{u',u}|^{2}
\right).
\end{equation}

In terms of the original quantities,
\begin{equation}
L_{u',u}
=
\left(\frac{D_u}{\gamma}\right)^3\left[I_{1,z,u'}
+
\frac{
I_{2,z,u'}
}{
3k-\widetilde{\omega}^{(3)}
+\iu\left(
\widetilde{\gamma}^{(3)}
+\widetilde{\gamma}^{(3)}_{\rm nr}
\right)
}\right],
\end{equation}
and
\begin{equation}
\mathcal{N}_{u',u}
=
9|L_{u',u}|^{2}
\left|
\bar{\chi}_{0}^{(3)}
\right|^{2}
+
\frac{
\left|
\Delta\bar{\chi}^{(3)}
\right|^{2}
}{2}
\left(
|L_{u',u}|^{2}+\frac{|D_u|^6|I_{1,z,u'}|^{2}}{\gamma^6}
\right).
\end{equation}

In particular, if the TH modes are not excited efficiently, $\mathcal{R}_{u',u}\to 0$, thus $B_{u',u}=D_{u',u}=0$. Then, 
\begin{equation}
{\rm CD}_{u',u}^{(3)}
\approx
\frac{
A_{u',u}\sin(4\phi)}{
1+C_{u',u}\cos(4\phi)
}.
\end{equation}

\newpage

\section{Numerical comparison with other cubic nonlinear materials}\label{sm:sec:materials} 

In this section, we compare the nonlinear response of resonant metasurfaces composed of different cubic nonlinear materials, including Ge, cubic diamond (C), Cu$_2$O, and CaF$_2$. For all materials, we employ the same metasurface geometry consisting of a suspended dielectric slab perforated by a square lattice of circular air holes, as shown in Fig.~\ref{fig:2}(b) of the main text. The lattice period $p$ and hole size parameter $\delta=d/p$ are selected for each material to obtain the desired pump dipolar resonance. The material parameters used in the numerical calculations and their literature sources are provided below.

For Ge, we use the refractive index $n(\omega)=4.10$ at the pump frequency and the complex refractive index $n(3\omega)+ik(3\omega)=5.19+0.53i$ at the TH frequency~\cite{sm:nunley2016optical}. The nonlinear anisotropy parameter is $3\chi_{1212}^{(3)}/\chi_{1111}^{(3)}=1.41e^{i\pi/10}$~\cite{sm:moss1990band}. The metasurface parameters are $p=680$~nm, $h=220$~nm, and $\delta=0.20$, corresponding to a pump dipolar resonance at approximately 146~THz.

For cubic diamond (C), we use the refractive indices $n(\omega)=2.39$ and $n(3\omega)=2.42$~\cite{sm:phillip1964diamond}. The nonlinear anisotropy parameter is $3\chi_{1212}^{(3)}/\chi_{1111}^{(3)}=1.30$~\cite{sm:moss1991semi}. The metasurface parameters are $p=725$~nm, $h=220$~nm, and $\delta=0.20$, corresponding to a pump dipolar resonance at approximately 232~THz.

For Cu$_2$O, we use $n(\omega)=2.63$ and $n(3\omega)+ik(3\omega)=3.45+0.12i$ at the corresponding pump and TH frequencies~\cite{sm:malerba2011absorption}. The nonlinear anisotropy parameter is $3\chi_{1212}^{(3)}/\chi_{1111}^{(3)}=0.27$~\cite{sm:frazer2014third}. The metasurface parameters are $p=785$~nm, $h=220$~nm, and $\delta=0.20$, corresponding to a pump dipolar resonance at approximately 200~THz.

For CaF$_2$, we use the refractive indices $n(\omega)=1.43$ and $n(3\omega)=1.44$~\cite{sm:malitson1963calcium}. The nonlinear anisotropy parameter is $3\chi_{1212}^{(3)}/\chi_{1111}^{(3)}=1.54$~\cite{sm:bombach1992third}. The metasurface parameters are $p=1385$~nm, $h=220$~nm, and $\delta=0.23$, corresponding to a pump dipolar resonance at approximately 202~THz.

\begin{figure}[H]
\centering
\includegraphics[width=0.85\linewidth]{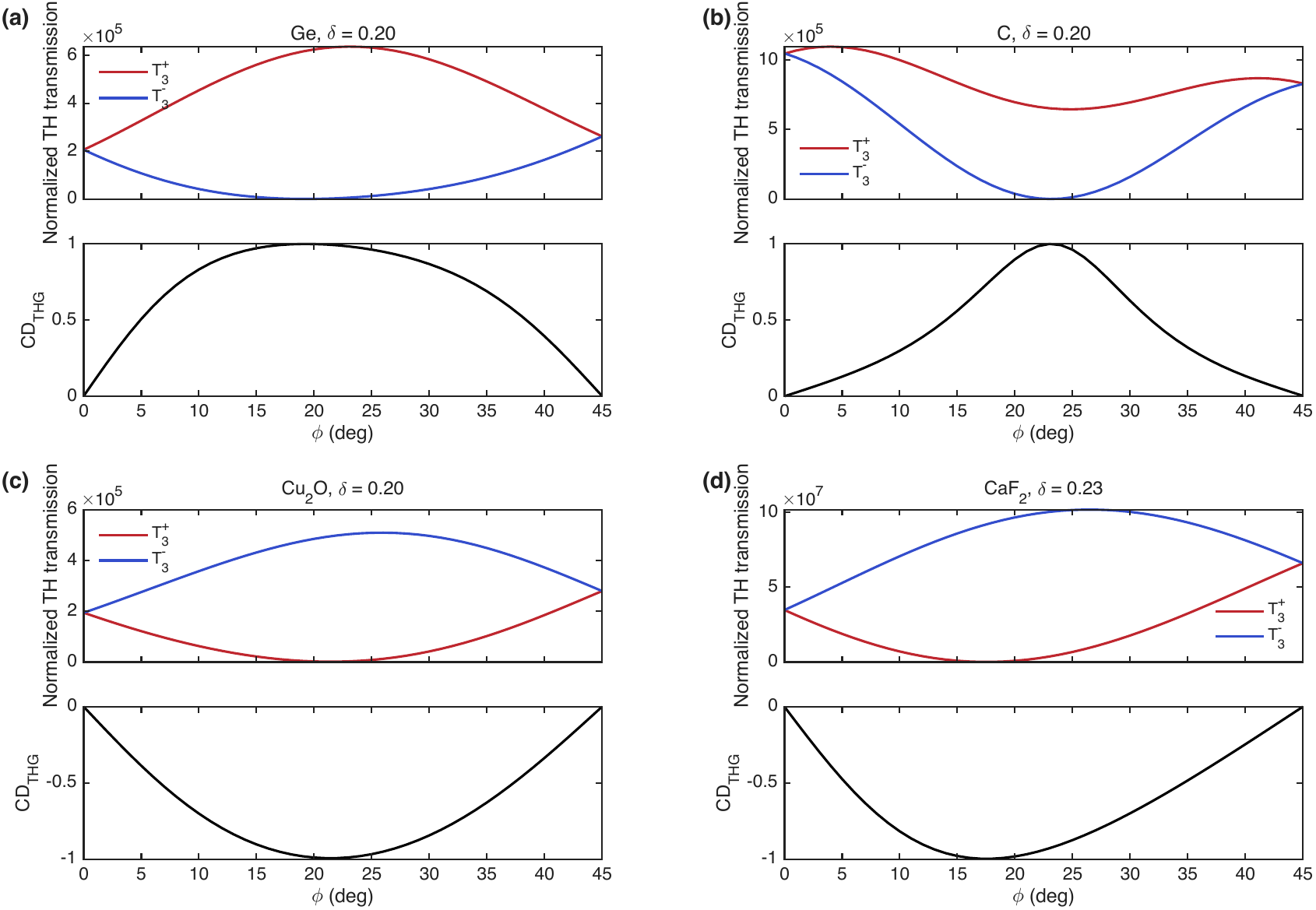}
\caption{
{\bf Nonlinear resonant response for different cubic nonlinear materials.}
(a--d) Normalized TH transmission $T_{+}^{(3)}$ and $T_{-}^{(3)}$ (upper panels) and the corresponding TH CD $CD^{(3)}$ (lower panels) vs. the relative angle $\phi$ between the metasurface and $\hat{\chi}^{(3)}$ in-plane axes for (a) Ge with $\delta=0.20$, (b) cubic diamond (C) with $\delta=0.20$, (c) Cu$_2$O with $\delta=0.20$, and (d) CaF$_2$ with $\delta=0.23$. The corresponding pump resonance frequencies are $146$~THz, $232$~THz, $200$~THz, and $202$~THz, respectively, with RCP ($\sigma=1$) and LCP ($\sigma=-1$) excitation. The TH transmission for each material is normalized to the respective TH transmission intensity of an unpatterned film of the same material and thickness.
}
\label{sm:fig:S6}
\end{figure}

Figure~\ref{sm:fig:S6} shows the corresponding nonlinear response as a function of the relative angle $\phi$ between the in-plane axes of the metasurface and the nonlinear susceptibility tensor $\hat{\chi}^{(3)}$. Following the normalization used in the main text, the TH transmission for each material is normalized to the respective TH transmission intensity of an unpatterned film of the same material and thickness.

As shown in Fig.~\ref{sm:fig:S6}, strongly helicity-selective THG is obtained for all four materials. At the corresponding optimal relative angles $\phi$, the TH signal for one input helicity is strongly suppressed, resulting in $|CD^{(3)}|>0.99$ in all cases. Specifically, the maximum $CD^{(3)}$ reaches 0.998 for Ge and cubic diamond, while values of $-0.994$ and $-0.997$ are obtained for Cu$_2$O and CaF$_2$, respectively. Despite the substantial differences in their linear optical constants and nonlinear anisotropy parameters, near-unity resonant TH circular dichroism can therefore be obtained for all considered cubic nonlinear materials.


\begin{thebibliography}{10}
\expandafter\ifx\csname url\endcsname\relax
  \def\url#1{\texttt{#1}}\fi
\expandafter\ifx\csname urlprefix\endcsname\relax\def\urlprefix{URL }\fi
\expandafter\ifx\csname href\endcsname\relax
  \def\href#1#2{#2} \def\path#1{#1}\fi

\bibitem{khaliq2023recent}
H.~S. Khaliq, A.~Nauman, J.-W. Lee, H.-R. Kim, Recent progress on plasmonic and
  dielectric chiral metasurfaces: fundamentals, design strategies, and
  implementation, Advanced Optical Materials 11~(16) (2023) 2300644.

\bibitem{fagiani2023modelling}
L.~Fagiani, M.~Gandolfi, L.~Carletti, C.~de~Angelis, J.~Osmond, M.~Bollani,
  Modelling and nanofabrication of chiral dielectric metasurfaces, Micro and
  Nano Engineering 19 (2023) 100187.

\bibitem{gorkunov2024rational}
M.~V. Gorkunov, A.~A. Antonov, Rational design of maximum chiral dielectric
  metasurfaces, in: All-Dielectric Nanophotonics, Elsevier, 2024, pp. 243--286.

\bibitem{hu2019high}
J.~Hu, M.~Lawrence, J.~A. Dionne, High quality factor dielectric metasurfaces
  for ultraviolet circular dichroism spectroscopy, Acs Photonics 7~(1) (2019)
  36--42.

\bibitem{li2025valley}
C.~Li, K.~Xing, W.~Zhai, L.~Sortino, A.~Tittl, I.~Aharonovich, M.~S. Fuhrer,
  K.~Watanabe, T.~Taniguchi, Q.~Ou, et~al., Valley optoelectronics based on
  meta-waveguide photodetectors, arXiv preprint arXiv:2503.19565 (2025).

\bibitem{ni2021giant}
J.~Ni, S.~Liu, G.~Hu, Y.~Hu, Z.~Lao, J.~Li, Q.~Zhang, D.~Wu, S.~Dong, J.~Chu,
  et~al., Giant helical dichroism of single chiral nanostructures with photonic
  orbital angular momentum, ACS nano 15~(2) (2021) 2893--2900.

\bibitem{wang2025resonant}
Y.~Wang, C.~Li, H.~Yu, S.~A. Maier, J.~Shi, H.~Ren, K.~Koshelev, Resonant
  helical dichroism in twisted dielectric metastructures, ACS nano 19~(35)
  (2025) 31894--31900.

\bibitem{solomon2018enantiospecific}
M.~L. Solomon, J.~Hu, M.~Lawrence, A.~Garc{\'\i}a-Etxarri, J.~A. Dionne,
  Enantiospecific optical enhancement of chiral sensing and separation with
  dielectric metasurfaces, Acs Photonics 6~(1) (2018) 43--49.

\bibitem{both2022nanophotonic}
S.~Both, M.~Schaferling, F.~Sterl, E.~A. Muljarov, H.~Giessen, T.~Weiss,
  Nanophotonic chiral sensing: how does it actually work?, ACS nano 16~(2)
  (2022) 2822--2832.

\bibitem{tong2026deciphering}
L.~Tong, F.~Xie, X.~Gao, Y.~Chen, S.~Ji, B.~Zhang, J.~Li, J.~Guo, F.~Liu,
  C.~Gao, et~al., Deciphering light transformation in chiral metasurface in
  real space and time by ultrafast electron microscopy, Light: Science \&
  Applications 15~(1) (2026) 70.

\bibitem{koshelev2023nonlinear}
K.~Koshelev, P.~Tonkaev, Y.~Kivshar, Nonlinear chiral metaphotonics: a
  perspective, Advanced Photonics 5~(6) (2023) 064001--064001.

\bibitem{koshelev2023resonant}
K.~Koshelev, Y.~Tang, Z.~Hu, I.~I. Kravchenko, G.~Li, Y.~Kivshar, Resonant
  chiral effects in nonlinear dielectric metasurfaces, ACS photonics 10~(1)
  (2023) 298--306.

\bibitem{tonkaev2025nonlinear}
P.~Tonkaev, Y.~Zhuang, D.~Kim, I.~Toftul, T.~Yamaguchi, K.~Li, J.~Huang,
  H.~Wang, T.~Tanaka, H.-G. Park, et~al., Nonlinear chiral response from
  linearly achiral membrane metasurfaces, Nano Letters 25~(47) (2025)
  16643--16649.

\bibitem{nechayev2019substrate}
S.~Nechayev, R.~Barczyk, U.~Mick, P.~Banzer, Substrate-induced chirality in an
  individual nanostructure, ACS Photonics 6~(8) (2019) 1876--1881.

\bibitem{gorkunov2025substrate}
M.~V. Gorkunov, A.~A. Antonov, A.~V. Mamonova, E.~A. Muljarov, Y.~Kivshar,
  Substrate-induced maximum optical chirality of planar dielectric structures,
  Advanced Optical Materials 13~(3) (2025) 2402133.

\bibitem{gryb2023two}
D.~Gryb, F.~J. Wendisch, A.~Aigner, T.~Golz, A.~Tittl, L.~de~S.~Menezes, S.~A.
  Maier, Two-dimensional chiral metasurfaces obtained by geometrically simple
  meta-atom rotations, Nano Letters 23~(19) (2023) 8891--8897.

\bibitem{toftul2024chiral}
I.~Toftul, P.~Tonkaev, K.~Koshelev, F.~Lai, Q.~Song, M.~Gorkunov, Y.~Kivshar,
  Chiral dichroism in resonant metasurfaces with monoclinic lattices, Physical
  Review Letters 133~(21) (2024) 216901.

\bibitem{sinev2025chirality}
I.~Sinev, F.~U. Richter, I.~Toftul, N.~Glebov, K.~Koshelev, Y.~Hwang, D.~G.
  Lancaster, Y.~Kivshar, H.~Altug, Chirality encoding in resonant metasurfaces
  governed by lattice symmetries, nature communications 16~(1) (2025) 6091.

\bibitem{papakostas2003optical}
A.~Papakostas, A.~Potts, D.~Bagnall, S.~Prosvirnin, H.~Coles, N.~Zheludev,
  Optical manifestations of planar chirality, Physical review letters 90~(10)
  (2003) 107404.

\bibitem{ren2012giant}
M.~Ren, E.~Plum, J.~Xu, N.~I. Zheludev, Giant nonlinear optical activity in a
  plasmonic metamaterial, Nature communications 3~(1) (2012) 833.

\bibitem{zhang2022chiral}
X.~Zhang, Y.~Liu, J.~Han, Y.~Kivshar, Q.~Song, Chiral emission from resonant
  metasurfaces, Science 377~(6611) (2022) 1215--1218.

\bibitem{wang2024enhanced}
Y.~Wang, C.~Qin, H.~Hu, J.~Liu, C.~Guan, Y.~Kivshar, K.~Koshelev, J.~Shi,
  Enhanced intrinsic chiroptical response of resonant metallic metasurfaces,
  Optics Letters 49~(18) (2024) 5288--5291.

\bibitem{tanaka2020chiral}
K.~Tanaka, D.~Arslan, S.~Fasold, M.~Steinert, J.~Sautter, M.~Falkner,
  T.~Pertsch, M.~Decker, I.~Staude, Chiral bilayer all-dielectric metasurfaces,
  ACS nano 14~(11) (2020) 15926--15935.

\bibitem{gromyko2024unidirectional}
D.~Gromyko, S.~An, S.~Gorelik, J.~Xu, L.~J. Lim, H.~Y.~L. Lee,
  F.~Tjiptoharsono, Z.-K. Tan, C.-W. Qiu, Z.~Dong, et~al., Unidirectional
  chiral emission via twisted bi-layer metasurfaces, Nature Communications
  15~(1) (2024) 9804.

\bibitem{wang2025chiral}
B.~Wang, T.~Zhu, Y.~Liu, H.~Yang, R.~Pan, J.~Li, Chiral resonant modes induced
  by intrinsic birefringence in lithium niobate metasurfaces, Physical Review
  Letters 134~(11) (2025) 113802.

\bibitem{ji2026intrinsically}
C.-Y. Ji, G.~Hu, Intrinsically chiral 1d planar photonic crystal slabs, Laser
  \& Photonics Reviews 20~(2) (2026) e01164.

\bibitem{frizyuk2021nonlinear}
K.~Frizyuk, E.~Melik-Gaykazyan, J.-H. Choi, M.~I. Petrov, H.-G. Park,
  Y.~Kivshar, Nonlinear circular dichroism in mie-resonant nanoparticle dimers,
  Nano Letters 21~(10) (2021) 4381--4387.

\bibitem{li2026meta}
B.~Li, Z.~Wang, Q.~Fang, N.~Lu, Z.~Han, W.~Chen, Y.~Jia, F.~Chen, Meta-atom
  rotation unlocks nonlinear optical chirality in lithium niobate metasurfaces,
  Laser \& Photonics Reviews (2026) e71170.

\bibitem{jangid2025chiral}
P.~Jangid, M.~A. Vincenti, L.~Carletti, A.~Rudenko, Y.~Kivshar, S.~Kruk, Chiral
  high-harmonic generation in metasurfaces, ACS photonics 12~(8) (2025)
  4342--4348.

\bibitem{heimig2026chiral}
C.~Heimig, A.~A. Antonov, D.~Gryb, T.~Possmayer, T.~Weber, M.~Hirler,
  J.~Biechteler, L.~Sortino, L.~de~S.~Menezes, S.~A. Maier, et~al., Chiral
  nonlinear polaritonics with van der waals metasurfaces, Science Advances
  12~(13) (2026) eaeb5631.

\bibitem{koshelev2024scattering}
K.~Koshelev, I.~Toftul, Y.~Hwang, Y.~Kivshar, Scattering matrix for chiral
  harmonic generation and frequency mixing in nonlinear metasurfaces, Journal
  of Optics 26~(5) (2024) 055003.

\bibitem{boroviks2023demonstration}
S.~Boroviks, A.~Kiselev, K.~Achouri, O.~J. Martin, Demonstration of a plasmonic
  nonlinear pseudodiode, Nano Letters 23~(8) (2023) 3362--3368.

\bibitem{kruk2022asymmetric}
S.~S. Kruk, L.~Wang, B.~Sain, Z.~Dong, J.~Yang, T.~Zentgraf, Y.~Kivshar,
  Asymmetric parametric generation of images with nonlinear dielectric
  metasurfaces, Nature Photonics 16~(8) (2022) 561--565.

\bibitem{liu2025dynamic}
Y.~Liu, C.~Meng, S.~I. Bozhevolnyi, F.~Ding, Dynamic and continuous control of
  second-harmonic chirality through lithium niobate nonlocal metasurface, arXiv
  preprint arXiv:2509.12745 (2025).

\bibitem{lai2025nonlinear}
F.~Lai, J.~Yin, I.~Toftul, H.~Barkaoui, H.~Deng, X.~Sha, M.~V. Gorkunov,
  Y.~Kivshar, Q.~Song, Nonlinear chiral light generation from resonant
  metasurfaces, Nature Communications 16~(1) (2025) 10686.

\bibitem{tian2026wavefront}
Y.~Tian, N.~Wang, Q.~Liu, S.~Xiao, T.~Liu, O.~J. Martin, Y.~Gu, Wavefront
  control and intensity modulation of third harmonic generation in nonlocal
  metasurfaces, ACS Photonics (2026).

\bibitem{gromyko2025enabling}
D.~Gromyko, J.~S. Loh, J.~Feng, C.-W. Qiu, L.~Wu, Enabling all-to-circular
  polarization up-conversion by nonlinear chiral metasurfaces with rotational
  symmetry, Physical Review Letters 134~(2) (2025) 023804.

\bibitem{chen2014symmetry}
S.~Chen, G.~Li, F.~Zeuner, W.~H. Wong, E.~Y.~B. Pun, T.~Zentgraf, K.~W. Cheah,
  S.~Zhang, Symmetry-selective third-harmonic generation from plasmonic
  metacrystals, Physical review letters 113~(3) (2014) 033901.

\bibitem{konishi2014polarization}
K.~Konishi, T.~Higuchi, J.~Li, J.~Larsson, S.~Ishii, M.~Kuwata-Gonokami,
  Polarization-controlled circular second-harmonic generation from metal hole
  arrays with threefold rotational symmetry, Physical review letters 112~(13)
  (2014) 135502.

\bibitem{achouri2023spatial}
K.~Achouri, V.~Tiukuvaara, O.~J. Martin, Spatial symmetries in nonlocal
  multipolar metasurfaces, Advanced Photonics 5~(4) (2023) 046001--046001.

\bibitem{nikitina2023nonlinear}
A.~Nikitina, A.~Nikolaeva, K.~Frizyuk, Nonlinear circular dichroism in achiral
  dielectric nanoparticles, Physical Review B 107~(4) (2023) L041405.

\bibitem{nikitina2024achiral}
A.~Nikitina, K.~Frizyuk, Achiral nanostructures: perturbative harmonic
  generation and dichroism under vortex and vector beams illumination, Advanced
  Optical Materials 12~(25) (2024) 2400732.

\bibitem{menshikov2026tailoring}
E.~Menshikov, C.~De~Angelis, K.~Frizyuk, Tailoring nonlinear circular dichroism
  in nanostructures and metasurfaces, Journal of the Optical Society of America
  B 43~(10) (2026) C32--C40.

\bibitem{toftul2023nonlinearity}
I.~Toftul, G.~Fedorovich, D.~Kislov, K.~Frizyuk, K.~Koshelev, Y.~Kivshar,
  M.~Petrov, Nonlinearity-induced optical torque, Physical Review Letters
  130~(24) (2023) 243802.

\bibitem{menshikov2025light}
E.~Menshikov, P.~Franceschini, K.~Frizyuk, I.~Fernandez-Corbaton, A.~Tognazzi,
  A.~C. Cino, D.~Garoli, M.~Petrov, D.~De~Ceglia, C.~De~Angelis, Light
  structuring via nonlinear total angular momentum addition with flat optics,
  Light: Science \& Applications 14~(1) (2025) 381.

\bibitem{menshikov2026near}
E.~Menshikov, M.~A. L{\'o}pez~S{\'a}nchez, P.~Franceschini, A.~Tognazzi,
  D.~De~Ceglia, K.~Frizyuk, C.~De~Angelis, Near-infrared polarimetric imaging
  with nonlinear flat-optics, Laser \& Photonics Reviews (2026) e03276.

\bibitem{guercio2026tensor}
G.~Guercio, A.~Gerini, K.~Frizyuk, M.~Morassi, A.~Lemaitre, C.~De~Angelis,
  L.~Carletti, G.~Leo, Tensor-driven geometric phase in nonlinear algaas
  metasurfaces, ACS Photonics (2026).

\bibitem{overvig2020selection}
A.~C. Overvig, S.~C. Malek, M.~J. Carter, S.~Shrestha, N.~Yu, Selection rules
  for quasibound states in the continuum, Physical Review B 102~(3) (2020)
  035434.

\bibitem{fan2003temporal}
S.~Fan, W.~Suh, J.~D. Joannopoulos, Temporal coupled-mode theory for the fano
  resonance in optical resonators, Journal of the Optical Society of America A
  20~(3) (2003) 569--572.

\bibitem{ruan2012temporal}
Z.~Ruan, S.~Fan, Temporal coupled-mode theory for light scattering by an
  arbitrarily shaped object supporting a single resonance, Physical Review A
  85~(4) (2012) 043828.

\bibitem{liang2011three}
Y.~Liang, C.~Peng, K.~Sakai, S.~Iwahashi, S.~Noda, Three-dimensional
  coupled-wave model for square-lattice photonic crystal lasers with transverse
  electric polarization: A general approach, Physical Review B 84~(19) (2011)
  195119.

\bibitem{franta2018determination}
D.~Franta, P.~Franta, J.~Voh{\'a}nka, M.~{\v{C}}erm{\'a}k, I.~Ohl{\'\i}dal,
  Determination of thicknesses and temperatures of crystalline silicon wafers
  from optical measurements in the far infrared region, Journal of Applied
  Physics 123~(18) (2018).

\bibitem{alexysong_inkstone}
A.~Y. Song, {Inkstone}: Efficient electromagnetic solver based on rigorous
  coupled-wave analysis, \href{https://github.com/alexysong/inkstone}{\texttt{https://github.com/alexysong/inkstone}}, accessed:
  2025-08-11 (2025).

\bibitem{moss1989dispersion}
D.~Moss, H.~M. van Driel, J.~E. Sipe, Dispersion in the anisotropy of optical
  third-harmonic generation in silicon, Optics letters 14~(1) (1989) 57--59.

\end{thebibliography}

\begin{thebibliography}{S14}%
\makeatletter
\providecommand \@ifxundefined [1]{%
 \@ifx{#1\undefined}
}%
\providecommand \@ifnum [1]{%
 \ifnum #1\expandafter \@firstoftwo
 \else \expandafter \@secondoftwo
 \fi
}%
\providecommand \@ifx [1]{%
 \ifx #1\expandafter \@firstoftwo
 \else \expandafter \@secondoftwo
 \fi
}%
\providecommand \natexlab [1]{#1}%
\providecommand \enquote  [1]{``#1''}%
\providecommand \bibnamefont  [1]{#1}%
\providecommand \bibfnamefont [1]{#1}%
\providecommand \citenamefont [1]{#1}%
\providecommand \href@noop [0]{\@secondoftwo}%
\providecommand \href [0]{\begingroup \@sanitize@url \@href}%
\providecommand \@href[1]{\@@startlink{#1}\@@href}%
\providecommand \@@href[1]{\endgroup#1\@@endlink}%
\providecommand \@sanitize@url [0]{\catcode `\\12\catcode `\$12\catcode
  `\&12\catcode `\#12\catcode `\^12\catcode `\_12\catcode `\%12\relax}%
\providecommand \@@startlink[1]{}%
\providecommand \@@endlink[0]{}%
\providecommand \url  [0]{\begingroup\@sanitize@url \@url }%
\providecommand \@url [1]{\endgroup\@href {#1}{\urlprefix }}%
\providecommand \urlprefix  [0]{URL }%
\providecommand \Eprint [0]{\href }%
\providecommand \doibase [0]{https://doi.org/}%
\providecommand \selectlanguage [0]{\@gobble}%
\providecommand \bibinfo  [0]{\@secondoftwo}%
\providecommand \bibfield  [0]{\@secondoftwo}%
\providecommand \translation [1]{[#1]}%
\providecommand \BibitemOpen [0]{}%
\providecommand \bibitemStop [0]{}%
\providecommand \bibitemNoStop [0]{.\EOS\space}%
\providecommand \EOS [0]{\spacefactor3000\relax}%
\providecommand \BibitemShut  [1]{\csname bibitem#1\endcsname}%
\let\auto@bib@innerbib\@empty
\SMbibitem{sm:liang2011three}%
  \BibitemOpen
  \bibfield  {author} {\bibinfo {author} {\bibfnamefont {Y.}~\bibnamefont
  {Liang}}, \bibinfo {author} {\bibfnamefont {C.}~\bibnamefont {Peng}},
  \bibinfo {author} {\bibfnamefont {K.}~\bibnamefont {Sakai}}, \bibinfo
  {author} {\bibfnamefont {S.}~\bibnamefont {Iwahashi}},\ and\ \bibinfo
  {author} {\bibfnamefont {S.}~\bibnamefont {Noda}},\ }\bibfield  {title}
  {\bibinfo {title} {Three-dimensional coupled-wave model for square-lattice
  photonic crystal lasers with transverse electric polarization: A general
  approach},\ }\href@noop {} {\bibfield  {journal} {\bibinfo  {journal}
  {Physical Review B}\ }\textbf {\bibinfo {volume} {84}},\ \bibinfo {pages}
  {195119} (\bibinfo {year} {2011})}\BibitemShut {NoStop}%
\SMbibitem{sm:ruan2012temporal}%
  \BibitemOpen
  \bibfield  {author} {\bibinfo {author} {\bibfnamefont {Z.}~\bibnamefont
  {Ruan}}\ and\ \bibinfo {author} {\bibfnamefont {S.}~\bibnamefont {Fan}},\
  }\bibfield  {title} {\bibinfo {title} {Temporal coupled-mode theory for light
  scattering by an arbitrarily shaped object supporting a single resonance},\
  }\href@noop {} {\bibfield  {journal} {\bibinfo  {journal} {Physical Review
  A}\ }\textbf {\bibinfo {volume} {85}},\ \bibinfo {pages} {043828} (\bibinfo
  {year} {2012})}\BibitemShut {NoStop}%
\SMbibitem{sm:fan2003temporal}%
  \BibitemOpen
  \bibfield  {author} {\bibinfo {author} {\bibfnamefont {S.}~\bibnamefont
  {Fan}}, \bibinfo {author} {\bibfnamefont {W.}~\bibnamefont {Suh}},\ and\
  \bibinfo {author} {\bibfnamefont {J.~D.}\ \bibnamefont {Joannopoulos}},\
  }\bibfield  {title} {\bibinfo {title} {Temporal coupled-mode theory for the
  fano resonance in optical resonators},\ }\href@noop {} {\bibfield  {journal}
  {\bibinfo  {journal} {Journal of the Optical Society of America A}\ }\textbf
  {\bibinfo {volume} {20}},\ \bibinfo {pages} {569} (\bibinfo {year}
  {2003})}\BibitemShut {NoStop}%
\SMbibitem{sm:menshikov2026tailoring}%
  \BibitemOpen
  \bibfield  {author} {\bibinfo {author} {\bibfnamefont {E.}~\bibnamefont
  {Menshikov}}, \bibinfo {author} {\bibfnamefont {C.}~\bibnamefont
  {De~Angelis}},\ and\ \bibinfo {author} {\bibfnamefont {K.}~\bibnamefont
  {Frizyuk}},\ }\bibfield  {title} {\bibinfo {title} {Tailoring nonlinear
  circular dichroism in nanostructures and metasurfaces},\ }\href@noop {}
  {\bibfield  {journal} {\bibinfo  {journal} {Journal of the Optical Society of
  America B}\ }\textbf {\bibinfo {volume} {43}},\ \bibinfo {pages} {C32}
  (\bibinfo {year} {2026})}\BibitemShut {NoStop}%
\SMbibitem{sm:konishi2014polarization}%
  \BibitemOpen
  \bibfield  {author} {\bibinfo {author} {\bibfnamefont {K.}~\bibnamefont
  {Konishi}}, \bibinfo {author} {\bibfnamefont {T.}~\bibnamefont {Higuchi}},
  \bibinfo {author} {\bibfnamefont {J.}~\bibnamefont {Li}}, \bibinfo {author}
  {\bibfnamefont {J.}~\bibnamefont {Larsson}}, \bibinfo {author} {\bibfnamefont
  {S.}~\bibnamefont {Ishii}},\ and\ \bibinfo {author} {\bibfnamefont
  {M.}~\bibnamefont {Kuwata-Gonokami}},\ }\bibfield  {title} {\bibinfo {title}
  {Polarization-controlled circular second-harmonic generation from metal hole
  arrays with threefold rotational symmetry},\ }\href@noop {} {\bibfield
  {journal} {\bibinfo  {journal} {Physical review letters}\ }\textbf {\bibinfo
  {volume} {112}},\ \bibinfo {pages} {135502} (\bibinfo {year}
  {2014})}\BibitemShut {NoStop}%
\SMbibitem{sm:koshelev2024scattering}%
  \BibitemOpen
  \bibfield  {author} {\bibinfo {author} {\bibfnamefont {K.}~\bibnamefont
  {Koshelev}}, \bibinfo {author} {\bibfnamefont {I.}~\bibnamefont {Toftul}},
  \bibinfo {author} {\bibfnamefont {Y.}~\bibnamefont {Hwang}},\ and\ \bibinfo
  {author} {\bibfnamefont {Y.}~\bibnamefont {Kivshar}},\ }\bibfield  {title}
  {\bibinfo {title} {Scattering matrix for chiral harmonic generation and
  frequency mixing in nonlinear metasurfaces},\ }\href@noop {} {\bibfield
  {journal} {\bibinfo  {journal} {Journal of Optics}\ }\textbf {\bibinfo
  {volume} {26}},\ \bibinfo {pages} {055003} (\bibinfo {year}
  {2024})}\BibitemShut {NoStop}%
\SMbibitem{sm:nunley2016optical}%
  \BibitemOpen
  \bibfield  {author} {\bibinfo {author} {\bibfnamefont {T.~N.}\ \bibnamefont
  {Nunley}}, \bibinfo {author} {\bibfnamefont {N.~S.}\ \bibnamefont
  {Fernando}}, \bibinfo {author} {\bibfnamefont {N.}~\bibnamefont
  {Samarasingha}}, \bibinfo {author} {\bibfnamefont {J.~M.}\ \bibnamefont
  {Moya}}, \bibinfo {author} {\bibfnamefont {C.~M.}\ \bibnamefont {Nelson}},
  \bibinfo {author} {\bibfnamefont {A.~A.}\ \bibnamefont {Medina}},\ and\
  \bibinfo {author} {\bibfnamefont {S.}~\bibnamefont {Zollner}},\ }\bibfield
  {title} {\bibinfo {title} {Optical constants of germanium and thermally grown
  germanium dioxide from 0.5 to 6.6 ev via a multisample ellipsometry
  investigation},\ }\href {https://doi.org/10.1116/1.4963075} {\bibfield
  {journal} {\bibinfo  {journal} {Journal of Vacuum Science \& Technology B}\
  }\textbf {\bibinfo {volume} {34}},\ \bibinfo {pages} {061205} (\bibinfo
  {year} {2016})}\BibitemShut {NoStop}%
\SMbibitem{sm:moss1990band}%
  \BibitemOpen
  \bibfield  {author} {\bibinfo {author} {\bibfnamefont {D.~J.}\ \bibnamefont
  {Moss}}, \bibinfo {author} {\bibfnamefont {E.}~\bibnamefont {Ghahramani}},
  \bibinfo {author} {\bibfnamefont {J.~E.}\ \bibnamefont {Sipe}},\ and\
  \bibinfo {author} {\bibfnamefont {H.~M.}\ \bibnamefont {van Driel}},\
  }\bibfield  {title} {\bibinfo {title} {Band-structure calculation of
  dispersion and anisotropy in $\chi^{(3)}$ for third-harmonic generation in
  si, ge, and gaas},\ }\href {https://doi.org/10.1103/PhysRevB.41.1542}
  {\bibfield  {journal} {\bibinfo  {journal} {Physical Review B}\ }\textbf
  {\bibinfo {volume} {41}},\ \bibinfo {pages} {1542} (\bibinfo {year}
  {1990})}\BibitemShut {NoStop}%
\SMbibitem{sm:phillip1964diamond}%
  \BibitemOpen
  \bibfield  {author} {\bibinfo {author} {\bibfnamefont {H.~R.}\ \bibnamefont
  {Phillip}}\ and\ \bibinfo {author} {\bibfnamefont {E.~A.}\ \bibnamefont
  {Taft}},\ }\bibfield  {title} {\bibinfo {title} {Kramers-kronig analysis of
  reflectance data for diamond},\ }\href
  {https://doi.org/10.1103/PhysRev.136.A1445} {\bibfield  {journal} {\bibinfo
  {journal} {Physical Review}\ }\textbf {\bibinfo {volume} {136}},\ \bibinfo
  {pages} {A1445} (\bibinfo {year} {1964})}\BibitemShut {NoStop}%
\SMbibitem{sm:moss1991semi}%
  \BibitemOpen
  \bibfield  {author} {\bibinfo {author} {\bibfnamefont {D.~J.}\ \bibnamefont
  {Moss}}, \bibinfo {author} {\bibfnamefont {E.}~\bibnamefont {Ghahramani}},\
  and\ \bibinfo {author} {\bibfnamefont {J.~E.}\ \bibnamefont {Sipe}},\
  }\bibfield  {title} {\bibinfo {title} {Semi-ab initio tight-binding
  band-structure calculations of $\chi^{(3)}(-3\omega;\omega,\omega,\omega)$ in
  c, si, ge, sic, bp, alp, alas, alsb, gap, gaas, gasb, inp, inas, and insb},\
  }\href {https://doi.org/10.1002/pssb.2221640228} {\bibfield  {journal}
  {\bibinfo  {journal} {physica status solidi (b)}\ }\textbf {\bibinfo {volume}
  {164}},\ \bibinfo {pages} {587} (\bibinfo {year} {1991})}\BibitemShut
  {NoStop}%
\SMbibitem{sm:malerba2011absorption}%
  \BibitemOpen
  \bibfield  {author} {\bibinfo {author} {\bibfnamefont {C.}~\bibnamefont
  {Malerba}}, \bibinfo {author} {\bibfnamefont {F.}~\bibnamefont {Biccari}},
  \bibinfo {author} {\bibfnamefont {C.~L.}\ \bibnamefont {Azanza~Ricardo}},
  \bibinfo {author} {\bibfnamefont {M.}~\bibnamefont {D'Incau}}, \bibinfo
  {author} {\bibfnamefont {P.}~\bibnamefont {Scardi}},\ and\ \bibinfo {author}
  {\bibfnamefont {A.}~\bibnamefont {Mittiga}},\ }\bibfield  {title} {\bibinfo
  {title} {Absorption coefficient of bulk and thin film cu$_2$o},\ }\href
  {https://doi.org/10.1016/j.solmat.2011.05.047} {\bibfield  {journal}
  {\bibinfo  {journal} {Solar Energy Materials and Solar Cells}\ }\textbf
  {\bibinfo {volume} {95}},\ \bibinfo {pages} {2848} (\bibinfo {year}
  {2011})}\BibitemShut {NoStop}%
\SMbibitem{sm:frazer2014third}%
  \BibitemOpen
  \bibfield  {author} {\bibinfo {author} {\bibfnamefont {L.}~\bibnamefont
  {Frazer}}, \bibinfo {author} {\bibfnamefont {R.~D.}\ \bibnamefont
  {Schaller}}, \bibinfo {author} {\bibfnamefont {K.~B.}\ \bibnamefont {Chang}},
  \bibinfo {author} {\bibfnamefont {J.~B.}\ \bibnamefont {Ketterson}},\ and\
  \bibinfo {author} {\bibfnamefont {K.~R.}\ \bibnamefont {Poeppelmeier}},\
  }\bibfield  {title} {\bibinfo {title} {Third-harmonic generation in cuprous
  oxide: efficiency determination},\ }\href
  {https://doi.org/10.1364/OL.39.000618} {\bibfield  {journal} {\bibinfo
  {journal} {Optics Letters}\ }\textbf {\bibinfo {volume} {39}},\ \bibinfo
  {pages} {618} (\bibinfo {year} {2014})}\BibitemShut {NoStop}%
\SMbibitem{sm:malitson1963calcium}%
  \BibitemOpen
  \bibfield  {author} {\bibinfo {author} {\bibfnamefont {I.~H.}\ \bibnamefont
  {Malitson}},\ }\bibfield  {title} {\bibinfo {title} {A redetermination of
  some optical properties of calcium fluoride},\ }\href
  {https://doi.org/10.1364/AO.2.001103} {\bibfield  {journal} {\bibinfo
  {journal} {Applied Optics}\ }\textbf {\bibinfo {volume} {2}},\ \bibinfo
  {pages} {1103} (\bibinfo {year} {1963})}\BibitemShut {NoStop}%
\SMbibitem{sm:bombach1992third}%
  \BibitemOpen
  \bibfield  {author} {\bibinfo {author} {\bibfnamefont {R.}~\bibnamefont
  {Bombach}}, \bibinfo {author} {\bibfnamefont {B.}~\bibnamefont
  {Hemmerling}},\ and\ \bibinfo {author} {\bibfnamefont {W.}~\bibnamefont
  {Hubschmid}},\ }\bibfield  {title} {\bibinfo {title} {Third-order
  anisotropies in cubic crystals},\ }\href
  {https://doi.org/10.1364/AO.31.000367} {\bibfield  {journal} {\bibinfo
  {journal} {Applied Optics}\ }\textbf {\bibinfo {volume} {31}},\ \bibinfo
  {pages} {367} (\bibinfo {year} {1992})}\BibitemShut {NoStop}%
\end{thebibliography}
\end{document}